\documentclass[12pt, a4paper,fleqn]{article}
\usepackage{float} 
\usepackage{tabularx}
\usepackage{tocbibind} 
\usepackage{caption}
\usepackage{setspace}
\usepackage{exscale, amsmath}
\usepackage{rawfonts,lscape}
\usepackage{amsfonts}
\usepackage{amssymb}
\usepackage{faktor}
\usepackage{polynom}
\usepackage[utf8]{inputenc}
\usepackage{dcolumn}
\usepackage{epstopdf}
\usepackage{graphicx}
\usepackage{calc}
\usepackage[hidelinks]{hyperref}
\usepackage[english]{babel}
\usepackage{mathdots}
\usepackage{multirow}
\usepackage{bigdelim}
\usepackage{tikz}
\usepackage{rotating,tabularx}
\usetikzlibrary{arrows}
\usetikzlibrary{fadings}
\usetikzlibrary{patterns}
\usepackage{setspace}
\usepackage{booktabs}
\usepackage{longtable}
\usepackage[font=small,labelfont=bf]{caption}
\usepackage{arydshln}
\usepackage{soul}  
\usepackage{siunitx} 

\newfont{\tabfont}{cmr7 at 7pt}
\begin{document}
\title{A Pipeline for Variable Selection and\\
False Discovery Rate Control With an\\
Application in Labor Economics}

\author{Sophie-Charlotte Klose\thanks{Department of Management, University of Duisburg-Essen, Germany,
		{\sl e-mail:} sophie-charlotte.klose@uni-due.de.}  \qquad Johannes Lederer\thanks{Department of Mathematics, Ruhr-University Bochum, Germany, {\sl e-mail:} johannes.lederer@rub.de. The authors sincerely thank Fang Xie for generously sharing the coding material used for implementing the aggregated FDR control into the knockoff-filter.}}

\date{\textbf{\today} }

\thispagestyle{empty} 	\maketitle

\newcommand{\jl}[1]{\color{red}#1\color{black}}

\begin{abstract}
	\noindent We introduce tools for \textit{controlled} variable selection to economists.   
	In particular, we apply a recently introduced aggregation scheme for false
	discovery rate (FDR) control to German administrative data to determine the parts of the individual employment histories that are relevant for the career outcomes of women. 
	Our results suggest that career outcomes can be predicted based on a small set of variables, such as daily earnings, wage increases in combination with a high level of education, employment status, and working experience.	
\end{abstract}



\newpage
\doublespacing

\section{Introduction}
Titles like ``Economics in the age of big data'' \cite{Einav2014}, ``Big data: new tricks for econometrics'' \cite{Varian2014} and ``Beyond prediction: using big data for policy problems'' \cite{Athey2017} highlight a shift in the scale of economic data towards \textit{Big Data}. 
The term Big Data refers to settings, where we observe information on a large number of units, or many pieces of information on each unit, or both, and often in a complex setting with several crosssections per unit.
Nowadays, empirical research in economics increasingly relies on newly available large-scale administrative data sets, which offer new possibilities for fitting models. 

In particular, Big Data applications bring the opportunity to fit high-resolution models.
Complex models arise naturally in Big Data, where the total number of measurements is large. 
In the extreme case, fitting sophisticated models can mean to deal with parameter spaces that are comparable ($p \approx n$) to the number of samples or even larger ($p \gg n$).
Parameter spaces that are too high-dimensional ($p \approx n$ or $p \gg n$) can not be examined with standard estimation methods and therefore rely on tools from high-dimensional statistics.
High-dimensional statistics comes into play whenever one fits complex models ($p$ is large).
In high-dimensional settings the number of variables is substantial, but there is often a sense that many of the variables are of minor importance or completely irrelevant.
The basic concept is to focus on the most relevant parts of the parameter space, which are identified by leveraging prior information. High-dimensional statistics complement classical estimators with so-called penalty or prior functions that formulate mathematically how ``likely'' or ``promising'' certain models are.
The data-driven calibrated tuning parameters weight the prior function and thus determine the degree of regularization, i.e. for example, the degree of feeding the bare measurements with additional information. 

The notion most commonly imposed on the prior functions in high-dimensional statistics is \textit{sparsity}. 
Sparsity means that the data generating process can be modelled accurately by using only a small number of variables even though the actual number of variables at hand is large.
The probably most discussed and prominent sparsity-inducing penalty function, such as in the Lasso literature \cite{Tibshirani1996}, is the $\ell_1$-prior, which we also focus on.
If the data generating process can be approximated by a sparse model, it makes sense to speak of variable selection and false discovery rate (FDR) control \cite{Benjamini1995}.
Variable selection means that we are interested in teasing apart the ``relevant variables'' (variables with non-zero-valued entries in the parameter vector) from the ``irrelevant variables'' (variables with zero-valued entries in the parameter vector).
FDR control means that we want to ensure that the expected fraction of false discoveries among all discoveries is not to high in order to guarantee that the selected variables are indeed the true ones.

Roughly speaking, sparsity is the main ingredient of variable selection, since we are often interested in finding a few of the important covariates, that is to extract the relevant variables from those which are scientifically not extremely useful for understanding the dependence between the response variable and the few truely relevant covariates.

The final goal is to draw inferences about the estimates $\hat{\boldsymbol{\beta}}$ of the selected variables. In most cases, it is impossible to recover the true subset $\mathcal{S}$ of variables with no error. Hence, we are naturally interested in procedures that keep the resulting error small. This is typically measured in terms of false positives and false negatives. It is also known as the type I and type II errors, respectively. False positives are falsely selected variables, and false negatives are falsely omitted variables. A suitable measure to evaluate the estimator's variable selection accuracy is, for example, the hamming distance, the sum of false positives and false negatives.\footnote{The hamming distance is defined as $\text{hd}:= \text{fp}+\text{fn}$, where $\text{fp}:=|\{j \in \{1,\dots,p\}:\hat{\beta}_j \neq 0 \ \text{and} \ \beta^*_j = 0 \}|$ and $\text{fn}:=|\{j \in \{1,\dots,p\}:\hat{\beta}_j=0 \ \text{and} \ \beta^*_j \neq 0 \}|$. } The smaller this number, the better the estimator's performance. 

How does variable selection and FDR control work?
In classic hypothesis testing we are commonly interested in controlling the Type I error at a certain significance level $\alpha$ by individually testing $p$ null-hypotheses of the form $\mathfrak{H}_{0,j}: \beta_j^* =0$, for $j \in \{1, \dots, p\}$. 
Extending the Type I error control to multiple testing, we are able to control the FDR. The FDR is the expected proportion of false discoveries and total discoveries. However, accurate hypothesis testing also requires that the (statistical) power, the proportion of correctly selected hypotheses and total number of true hypotheses, is large. The power is typically measured in terms of true positives. True positives are truely selected variables. Accordingly, the number of true positives is tp$:=|\{j \in \{1,\dots,p\}:\hat{\beta}_j \neq 0 \ \text{and} \ \beta^*_j \neq 0 \}|$.
 
Therefore, in addition to the FDR, we also control the \textit{aggregated} false discovery rate (AFDR) recently proposed by \cite{Xie2019}.
It maximizes the power while simultaneously guaranteeing FDR control.
This aggregation scheme is an improvement on the original FDR and has the advantage of having the same theoretical guarantees as the FDR. Consequently, we can use the same methods as for the FDR to control the AFDR. 

Indeed, we want to find as many variables as possible while at the same time not having too many false positives. We focus here on controlling the (A)FDR with the knockoff filter \cite{Barber2015}, a data-driven procedure. There is a variety of other procedures to control the (A)FDR, but the generality and flexibility of the knockoff-filter  makes it so worthwhile to introduce and prepare this procedure to economists. The knockoffs are designed to mimic the correlation structure of the design matrix $X$, in a way that allows for accurate (A)FDR control. The knockoff filter can be adapted for a broad class of models and for a variety of test statistics \cite{Candes2016,Romano2019}. Moreover, the knockoff procedure works under \textit{any} fixed design matrix $X$ and does not require the strong assumptions for the covariates, usually required for most variable selection techniques in high-dimensional statistics \cite{Zhao2006}. In fact, high-dimensional tools often have to limit the correlations between the covariates. Typically the irrepresentability condition is also required which limits the correlations between the ``relevant'' and ``irrelevant'' variables.

Why is variable selection and (A)FDR control for economists of interest? First of all, variable selection and (A)FDR control answer a simple, but very fruitful, question: which variables does an outcome of interest depend upon? Economists often want to know which demographic or socioeconomic variables affect future economic outcomes such as income. Answering this question is not easy, as economic data sources become more and more detailed, providing a flood of potential explanatory variables, often knowing full well that the outcome of interest only depends on a small fraction of it. Especially in administrative data, the main workhorse in labor economics, there is due to the longitudinal nature a deluge of explanatory variables possibly interacting in many different ways. So far, many economic papers have shied away from including many explanatory variables. They subconsciously assumed that the data generating process is characterized by only a few variables that matter for the outcome of interest. Important covariates are included based on economic reasoning. As consequence, relevant explanatory variables which enter the model with complex and very flexible functional form are usually not accounted for. Even though economists may believe that an economic outcome depends on a small set of variables, they have a priori little or no clue about which ones are relevant. Therefore, modern high-dimensional tools aimed to \textit{controlled} variable selection can help economists to tease apart the relevant from irrelevant variables and to achieve valid inference. 

So far, in the empirical research, economists limit the number of explanatory variables by hand, rather then choosing them in a data-driven manner. Only for \textit{prediction} tasks we already have many insights from the economic literature on data-driven techniques. In recent years, scholars adopted machine learning (ML) tools in economics. Indeed, most recent influential reviews of ML methods aimed to economists (for example \cite{Mullainathan2017,Athey2019}) introduce ML as a powerful tool to solve problems around prediction. There have been successful applications of the prediction methodology to policy problems, where ML tools have been embeeded in the context of economic decision-making \cite{Kleinberg2015,Kleinberg2018}. However, in the economic literature, there is less knowledge about variable selection and certainly not about (A)FDR control. Variable selection with (A)FDR control solves problems around parameter estimation, the main goal in applied economics. The (A)FDR control is a new framework of controlled variable selection. It achieves correct inference in such a broad setting by constructing so-called knockoff variables which serve as a kind of control group for the covariates. Therefore, we prepare this framework for economists and show that we can gain new insights for the empirical work. 
 
In this paper, we demonstrate the potentials of model-X (MX) knockoffs \cite{Candes2016} in the field of economics based on an empirical application towards the labor market. 
MX knockoffs is a new data-driven tool that allows to link a large number of potential covariates to an outcome of interest in a nonlinear fashion. 
It identifies a subset of important variables from a large set while controlling the (A)FDR.

In particular, we are interested in which variables from individual employment and wage histories affect future professional careers of women.
Due to the longitudinal nature of the administrative employment records there is complex information on individual employment histories (for example types of employment, wages, skill level, occupation, age, professional experience) for the entire elapsed working live possibly interacting in many different ways.
Despite of having a large set of potential covariates, we assume that the binary career outcome only depends on a small fraction of it.
The main challenge is to search for the few truely important variables which are linked to the response in a nonlinear fashion.
We use a binary choice model within the class of generalized linear models and apply $\ell_1$ penalized logistic regression.
To achieve valid inference in our setting, we apply the MX knockoffs to a data set from the Sample of Integrated Employment Biographies (SIAB).
Correct inference in our controlled variable selection problem means that we effectively control the (A)FDR even in logistic regression with a large set of covariates.
Thus, we can be pretty confident that the MX knockoff procedure correctly teases apart important from irrelevant variables while maximizing power and guaranteeing FDR control.

It is well known from the economic literature that a higher educational level is associated with better career outcomes in terms of higher earnings. There is a huge economic literature showing the positive labor market returns of a high educational attainment (for example \cite{Boeckerman2019,Card1999,Heckman2018}). 
There is also some indication in the literature that the occupational choice is related to the career outcomes \cite{Almquist1970}, but there seems to be no extensive analysis of which of the several hundred explanatory variables from individual employment and wage histories are truely associated with future professional careers of women.

Our results suggest, first, that individual employment histories have predictive power with regard to professional careers and second, that only a relatively small subset of information gained from employment records is crucial.

The remainder of this paper is organized as follows. Section \ref{Method} introduces the model set-up and the methodology. Section \ref{Data} describes the data, after which Section \ref{Results} presents the main empirical results. The final section concludes the paper.


\newcommand{\design}{\ensuremath{X}}
\newcommand{\timevar}{\ensuremath{t}} 
\newcommand{\designA}{\design[\timevar]}
\newcommand{\designa}{\ensuremath{\boldsymbol{x}}}
\newcommand{\designB}{\tilde{\design}}
\newcommand{\outcome}{\ensuremath{\boldsymbol{y}}}
\newcommand{\outcomeA}{\outcome[\timevar]}
\newcommand{\target}{\ensuremath{\boldsymbol{\beta}^*}}

\section{Methodology}\label{Method}
In our empirical application, we use a large-scale model that links a large set of potential explanatory variables to the response variable  in a nonlinear fashion. We want to discover which variables from individual employment histories are truely associated with future professional careers of women. The data consist of vector-valued samples, each of them describing the employment and wage situation of women over the past five years. The design matrix $X \in \mathbb{R}^{79\,782 \times 327}$ is prepocessed and scaled such that it follows a multivariate normal distribution. The outcome variable $\boldsymbol{y} \in \mathbb{R}^{79\,782}$ is an indicator that takes the value $1$ if the woman makes a professional career in 2010, otherwise 0.

\subsection{Model and assumptions}

We consider data in form of a real-valued deterministic design matrix $X \in \mathbb{R}^{n \times p}$ and a binary response $\boldsymbol{y} \in \mathbb{R}^n$. We denote the rows of $X$ by $\boldsymbol{x}_1, \dots,\boldsymbol{x}_n \in \mathbb{R}^p$ and the columns of $X$ by $\boldsymbol{x}^1, \dots,\boldsymbol{x}^p \in \mathbb{R}^n$ . The design matrix $X$ is assumed to be scaled such that $(X^{\top}X)_{jj}=n$ for $j \in \{1, \dots, p\}$.

We define the vector of residuals as $\boldsymbol{u} =(u_1, \dots, u_n)^{\top}$ with entries $u_i= y_i - \mathbb{P}(y_i=1 | \boldsymbol{x}_i)$ for $i \in [n]$. The vector $\boldsymbol{u}$ is random noise with mean zero, i.e. $\mathbb{E}(y_i - \mathbb{P}(y_i=1|\boldsymbol{x}_i)) = 0$ . Further, we assume $\mathbb{E}(u_i|\boldsymbol{x}_i) =0$ and $\mathbb{E}(u_i u_j) =0 \quad \forall i \neq j$, i.e. our models contain only exogenous variables and the error terms are assumed to be independent and identically distributed (i.i.d.).  

The design matrix $\design$ and the response vector $\outcome$ are linked to standard logistic regression model
\begin{align}\label{eq.1}
\mathbb{P}(y_i=1|\boldsymbol{x}_i) = \frac{\exp(\boldsymbol{x}_i^{\top} \ \boldsymbol{\beta}^*)}{1+ \exp(\boldsymbol{x}_i^{\top}  \ \boldsymbol{\beta}^*)} \quad (i=1,\dots,n),
\end{align}
where $\boldsymbol{\beta}^* \in \mathbb{R}^p$ is the unknown regression vector.

We assume known that our data generating process is \textit{sparse}, that is, only a small and a priori unknown number of predictors is relevant for predicting the careers of women ($|\{j:\beta_j\neq 0 \}|\ll \min\{n,p\}$). 
Sparsity can be motivated on economic grounds in situations where a researcher believes that the economic outcome can be modeled accurately by using only a small number of variables (relative to the sample size) but is unsure about the identity of the relevant variables. 
Traditionally in the empirical literature economists assumed that the model of interest is characterized by a small number of variables, and they limited the number of explanatory variables by hand, rather than choosing them in a data-driven manner. 
However, recent advances in high-dimensional modeling have prompted economists to take advantage of the recent big data revolution to deal with large dimensional data sets, in a way that maintains the interpretability of economic models by searching for the truely important variables (for example \cite{Belloni2011a,Fan2011}). 
The \textit{sparsity} assumption imposed there allows the effective use of a large set of covariates while at the same time maintaining the spirit of parsimonious models in the economic discipline.
For this reason, we follow this strand of literature on the sparse modeling of economic processes.

\subsection{Estimation}
We use penalized techniques to solve the logistic regression and choose the $\ell_1$ penalty throughout our estimations.
The $\ell_1$ penalty is suitable for sparsity because it  forms a square constraint region for the parameter vector, and the least-squares contours are likely to intercept the constraint region at the extremes, such that certain coordinates of the parameter vector are set to zero. 
Moreover the $\ell_1$ penalty is a convex function, which makes it convenient to optimize over.

The goal is to estimate the support set $\mathcal{S} = \text{supp}(\target)$ for the model in eq. \ref{eq.1}  with the family of regularized estimators
\begin{align}\label{eq. 2}
\hat{\boldsymbol{\beta}}[r,X,\outcome] \in \underset{\target \in \mathbb{R}^p}{\arg \min}  \{ \mathcal{L}[\target,X,\outcome]+ rh[\target]  \},
\end{align}
where $\mathcal{L}[\target,X , \outcome] =\sum_{i=1}^{n} (\log(1+\exp(\designa_i^{\top} \ \target ))- y_i \designa_i^{\top} \ \target) /n $ is the negative log-likelihood function, $r \in [0,\infty)$ is a tuning parameter and $h:\mathbb{R}^p \to [0,\infty]$ is a prior function which we specify as sparsity inducing prior function
\begin{align*}
h[\target]:= ||\boldsymbol{\beta}^*||_1,
\end{align*}
where $|\cdot|$ denotes the absolute value and $||\boldsymbol{\beta}^*||_1:=  \sum_{j=1}^p|\beta_j^*|$ is the $\ell_1$-prior.

Next, to avoid an unwanted overall shrinkage of the estimates imposed by the $l_1$ penalty, we refit the regularized estimators $\hat{\boldsymbol{\beta}}$ with subsequent logistic regression estimation on the support $\hat{\mathcal{S}}:=\text{supp}(\hat{\boldsymbol{\beta}})$:
\begin{align}\label{eq. 3}
\hat{\boldsymbol{\beta}}_{\text{refit}}[X_{\hat{\mathcal{S}}},\outcome] \in \underset{\substack{\target \in \mathbb{R}^p \\ \text{supp}[\target] \subset \hat{\mathcal{S}}}}{\arg \min}  \mathcal{L}[\target,X_{\hat{\mathcal{S}}},\outcome].
\end{align}

\subsection{Inference}
In this paper, we focus on controlling the aggregated false discovery rate (AFDR) \cite{Xie2019}, which we can define as follows: letting $\mathcal{S}^* := \{j \in \{1, \dots,p\}: \beta_j^* \neq 0 \}$ be the true support set and $\hat{\mathcal{S}}_q[[X,\outcome], k]$ an estimate of $\mathcal{S}^*$ operating on data $[X,\outcome]$. Then, the AFDR at target level $q = \sum_{i=1}^k q_i \in [0,1]$ is
\begin{align}\label{eq. 4}
\text{AFDR} := \mathbb{E} \Bigg [ \frac{| \hat{\mathcal{S}}_q[[X,\outcome], k] \backslash  \mathcal{S}^*|}{| \hat{\mathcal{S}}_q[[X,\outcome], k]| \vee 1} \Bigg ] \le q \quad \text{with} \quad \mathcal{S}_q[[X,\outcome], k]:= \cup_{i=1}^k \mathcal{S}_q[X,\outcome],
\end{align}
where $|\cdot|$ denotes the cardinality of a set. The AFDR scheme is nothing else as applying the FDR control method \cite{Benjamini1995} $k$ times with specific target levels $q_1, \dots, q_k$ and combining the results by taking the union. Consequently, for $k=1$, we consider the original FDR control scheme. Because the AFDR scheme is equipped with the same guarantees as the original FDR control method, the same procedures that are used to control the FDR can here be applied.\footnote{For the proof see \cite{Xie2019}.}

Next, we give a quick introduction to the model-X knockoff filter which we use in our empirical application to control the AFDR. The model-X (MX) knockoff filter is a method for high-dimensional controlled variable selection in any class of generalized linear models (GLM) \cite{Candes2016}. It extends the knockoff procedure that was originally designed for controlling the FDR in low-dimensional linear models \cite{Barber2015}. The key ingredient of the knockoff filter are the generated knockoff copies $\designB \in \mathbb{R}^{n \times p}$ for the design matrix $X$, which mimic the correlation structures between the variables and therefore serve as a control group for them to ensure that not too many irrelevant variables are selected. The final goal is to perform FDR control on the specific statistics based on both $X$ and $\designB$.

Denote $X = (\designa_1, \dots,\designa_n)^{\top}$ and $\designB=(\tilde{\designa}_1, \dots, \tilde{\designa}_n)^{\top}$. In this paper, we generate MX knockoffs $\designB$ from a multivariate normal distribution obeying
\begin{align}
\designB | X \stackrel{d}{\sim} N_p(\boldsymbol{\mu}, V),
\end{align}
where we assume $X \stackrel{d}{\sim} N_p(\boldsymbol{0}, \Sigma)$ with $\Sigma \in \mathbb{R}^{p \times p}$ being positive definite and $\boldsymbol{\mu}$ and $V$ satisfy
\begin{align}
\boldsymbol{\mu} &:= X - X \Sigma^{-1} \ \text{diag}\{\boldsymbol{s}\}, \\
V &:=2\text{diag}\{\boldsymbol{s}\} - \text{diag}\{\boldsymbol{s}\}\Sigma^{-1}\text{diag}\{\boldsymbol{s}\}, 
\end{align}
with $\boldsymbol{s} \in \mathbb{R}^p$ making $V$ positive definite. This way to construct knockoffs is most common in the literature (for example \cite{Barber2019,Candes2016,Barber2015,Xie2019}).

We consider the following penalized estimator for logistic regression which is augmented by the knockoffs
\begin{align}\label{eq. 5}
\hat{\boldsymbol{\beta}}[r,X,\designB,\outcome] \in \underset{\target \in \mathbb{R}^{2p}}{\arg \min}  \{\mathcal{L}[\target,[X \ \designB], \outcome]  + rh[\target]\},
\end{align}
where $\mathcal{L}[\target,[X \ \designB], \outcome] =\sum_{i=1}^{n} (\log(1+\exp([\designa \ \tilde{\designa}]_i^{\top} \ \target ))- y_i [\designa \ \tilde{\designa}]_i^{\top} \ \target) /n $ is the negative log-likelihood function, $[X \ \designB] \in \mathbb{R}^{n \times 2p}$ is an augmented matrix, $r \in [0,\infty)$ is a tuning parameter, and $h:\mathbb{R}^{2p} \to [0,\infty]$ is a prior function which we specify as sparsity inducing prior function
\begin{align*}
	h[\target]:= \sum_{j=1}^{2p} |\beta^*_j|.
\end{align*}
We use Cross-Validation (CV) and a recently introduced novel calibration scheme~\cite{Li2019} for $\ell_1$-penalized logistic regression for calibrating the tuning parameter in equation \ref{eq. 5}. 
The latter calibration scheme is based on simple tests along the tuning parameter path and is equipped with finite sample guarantees for feature selection. 

We consider here the Lasso Signed Max (LSM) \cite{Barber2015} and the Lasso coefficient-difference (LCD) \cite{Candes2016} as knockoff statistics. The LSM statistic denotes the maximum penalty coefficients of each variable entering in the model in \ref{eq. 5}. The LCD statistic 
denotes the difference between the absolute values of the Lasso coefficients of the original variables and the knockoff copies. Denote the LSM and LCD statistics by $(Z_1,\dots,Z_p, \tilde{Z}_1,\dots, \tilde{Z}_p)$, that is,
\begin{align*}
Z_j[X, \outcome]&:= \sup\{r:\hat{\beta}_j[r,X,\designB,\outcome] \neq 0\},   \quad \quad   \text{and} \quad          Z_j[X, \outcome]:= |\hat{\beta}_j[r,X,\designB,\outcome]|,\\
\tilde{Z}_j[X, \outcome]&:=\sup\{r: \hat{\beta}_{p+j}[r,X,\designB,\outcome] \neq 0\}, \quad   \text{and} \quad  \tilde{Z}_j[X, \outcome]:= |\hat{\beta}_{p+j}[r,X,\designB,\outcome]|,
\end{align*}
for $j \in \{1,\dots, p\}$. Then the LSM and LCD statistic vectors $W:=(W_1,\dots,W_p)^{\top}$ can be defined by
\begin{align*}
W_j[X, \outcome] := \max\{Z_j, \tilde{Z_j}\} \cdot \text{sign}(Z_j - \tilde{Z}_j),     \quad \text{and} \quad    W_j[X, \outcome] := Z_j[X, \outcome] - \tilde{Z}_j[X, \outcome],
\end{align*}
for $j \in \{1, \dots, p \}$.

Then, the data-dependent thresholds of the knockoff and knockoff$^+$ filter methods (two types of knockoff procedures) for a given FDR target $q\in [0,1] $ are defined by
\begin{align*}
T_q[X, \outcome] &:= \min \Big\{ t \in \mathcal{W}: \frac{\# \{j:W_j[X, \outcome] \le -t \}}{\# \{j:W_j[X, \outcome] \ge t \} } \le q \Big\}, \quad \quad (\text{knockoff}) \\
T_q^+[X, \outcome] &:= \min \Big\{ t \in \mathcal{W}: \frac{ 1 + \# \{j:W_j[X, \outcome] \le -t \}}{\# \{j:W_j[X, \outcome] \ge t \} \vee 1 } \le q \Big\}, \quad \quad (\text{knockoff}^+).
\end{align*}
Thus, the corresponding estimated support recovery sets are defined as
\begin{align*}
\hat{\mathcal{S}}_q[X, \outcome] &:= \{j: W_j[X, \outcome] \ge T_q[X, \outcome]   \}, \\
\hat{\mathcal{S}}_q^+[X, \outcome] &:= \{j: W_j[X, \outcome] \ge T_q^+[X, \outcome]   \}.
\end{align*}
The estimated support $\hat{\mathcal{S}}_q^+[X, \outcome]$ obtained by the knockoff$^+$ procedure satisfies the inquality in \ref{eq. 4}, whereas $\hat{\mathcal{S}}_q[X, \outcome]$ satisfies this theoretical bound for an approximated FDR which is less or equal to the FDR.

We use the (A)FDR control methods for variable selection in our highdimensional empirical application. 
To apply (A)FDR control to our data, we use the knockoff filter for logistic regression.
In line with the recommendations in~\cite{Xie2019} we simulate $k=3$ independent knockoffs $\tilde{\boldsymbol{X}}^1,\tilde{\boldsymbol{X}}^2, \tilde{\boldsymbol{X}}^3$ from a Gaussian distribution for the AFDR control method. 
As target FDR, we choose $q=0.1$.

\section{Data}\label{Data}
Our analysis draws on employment records from German administrative data. 
In particular, we use the Sample of Integrated Employment Biographies (SIAB).\footnote{The data basis of this project is the Scientific Use File (SUF) of the SIAB (version SIAB-Regionalfile 1975 - 2014 \cite{Ganzer2017}). The data was assessed via a guest stay at the Research Data Centre (FDZ) of the German Federal Employment Agency at the Institute for Employment Research (IAB) and then via controlled data processing at the FDZ.} 
It is provided by the Institut für Arbeitsmarkt- und Berufsforschung (IAB) in Nuremberg, and draws a 2 \% random sample from the population of Integrated Employment Biographies (IEB). 
The source of the data set comes from the registration procedure for social insurance. 
The SIAB has been used in a number of studies on career and wage outcomes of women (for example \cite{Adda2017,Ejrnaes2013}).
  
The sample of the IEB considered here covers all employees in Germany that contributed to the social security system between 1975 and 2014 or receiving transfer payments from the labor agency, or being registered job seekers. Civil servants, including teachers and self-employed, are not included in the sample. The sample provides detailed daily information on, for example, earnings, occupations, employment, tansitions in and out of work, periods of unemployment, unemployment and welfare benefits as well as basic demographics. An important advantage of the data set is that, because of the large sample size and the longitudinal nature of the administrative data, employment and wage histories are measured precisely.

\subsubsection*{Empirical data preparation}
Our goal is to discover which variables from employment records and basic demographics are truely associated with professional careers of women.
Due to methodological reasons, we concentrate on one particular year: 2010.\footnote{The proposed FDR pipeline requires the design matrix X to have independent and identically distributed rows.} As robustness check, we do the same analysis for two further years: 2009 and 2011. 
We focus on West-Germany and restrict our sample to birth cohorts of women between 1965 and 1975. 
The birth cohort restriction corresponds to the age groups between 35 and 45 years. 
Before age 35 many women still climb the career ladder, and have not yet reached the final top salary, while few women reach their career peak at an age older than 45. 
Because we do not predict when women will reach their professional career for the first time, we also leave women who already had a professional career before 2010 in the sample. 
Finally, we drop those women who do not have any employment, benefit, or job search record in the data either one or two years before the prediction year 2010. 
The idea behind this restriction is that we want to avoid having women in the sample who have emigrated from Germany or have died, and thus cannot have a career recorded in our data. Overall, the final sample includes middle-aged West German women with and without German citizenship, and with some recent attachement to the labor market. The sample consists of $N_1= 1\,567$ women with a career in 2010 and $N_2=78\,214$ women with no career in 2010.

To discover which variables from employment records and basic demographics affect future professional careers in 2010, we generate a plethora of separate variables from our SIAB data to capture the available information on the wage and (un-)employment histories in a very detailed and flexible way.
Especially the longitudinal nature of the administrative employment data allows to capture complex information on individual employment and earning histories (for example types of employment, wages, skill level) 
for the entire elapsed working live possibly interacting in many different ways.
To accurately record time-varying information of the elapsed (un-)employment history, we consider five lags for each of the baseline variables. 
For time-constant information (for example education) we only consider one lag.
Consequently, we use information from data spells until 2009.
Employment records from 2010 cannot be used for our analysis because of endogeneity issues.
Consider a woman making a professional career in November 2010. 
She has most likely find out about her upcoming promotion in contract negotiations with her employer a few months earlier, and her knowledge of the career peak may induce her to work less.

We consider 5 sociodemographic and 107 (un-)employment and wage history baseline predictor variables. 
The reason for including also sociodemographic variables that do not explicitly account for the employment and wage histories is that they are strongly associated with professional career outcomes. 
For instance, highly educated women are more likely to achieve a professional career than low educated women. 
This means, the skill level might be a powerful predictor for a professional career.

Finally, to select the main effects that are truely associated with professional careers, it is important to allow for a wide range of plausible interactions between the baseline variables. 
Thus, we interact the lagged terms of the associated baseline predictors with each other in many different ways, so that we end up with 327 sociodemographic, and employment and wage history variables. 
A full list of the included predictor variables can be found in the appendix. We only consider those interactions that make sense from an economic point of view, and that do not lead to perfect multicollinearity, that is $X^{\top}X$ is invertible. 
Moreover, our interacted time-varying variables come from the same time period because we want to avoid modeling complex dynamics between different predictors for the ease of the interpretation later on. 
We also interact time-constant variables with time-varying variables. Further, we assume that some of the predictor variables can be expressed as a linear combination of a number of basis functions of the original measured variables. 
Indeed, our two wage profile predictor variables can be expressed through the group of the wage variables. 
We also include non-linear effects for women's age and working experience by using second-order polynomials.

In our empirical application, we consider a large-scale design matrix $X \in \mathbb{R}^{79\,782 \times 327}$. 
Thus, we consider a setting in which both the sample size and the feature size are large, but the feature size is much smaller than the sample size ($n,p \gg 1, p \ll n$). 
Although we do not consider the case where $n,p \gg 1, p \approx n$ or $p \gg n$, and high-dimensional statistics become indispensable, our case ($p$ is large) is the typical scope of high-dimensional statistics.

The outcome variable $y_i$  is an indicator that takes the value 1 if the woman makes a professional career in 2010, otherwise 0. 
Making a professional career means, according to our definition, to achieve a high wage level. A difficulty of the SIAB data is that the wage variable is censored from above to ensure that high earners cannot be identified. 
Therefore, we define the binary career outcome variable based on the social security contribution ceiling imposed on the SIAB data\footnote{Note that the social security contribution ceiling varies for each year and between West- and East-Germnay.}: whenever the wage variable coincides with the the social security contribution ceiling, the career dummy takes the value 1, and 0 otherwise. 

\subsubsection*{Descriptives}
Table \ref{Tab:1} in the appendix provides detailed descriptive statistics of all baseline variables included in the analysis. 
Among the 112 baseline predictors,  32 are continuous (like for example age in years) and 80 are binary (like for example part-time employed). 
The predictors collect information on age, education, non-German citizenship, employment, part-time employment, marginal employment, daily earnings, different wage profiles, change of establishment, wage growth, experience, unemployment benefits, welfare benefits, registered job search and occupations. 

 \begin{table*}[h!]
	\begin{center}
		\small
		\caption{\small Descriptives}\label{table3.1}
		\begin{tabular}{p{4.5cm} >{\raggedleft}p{1.2cm} >{\raggedleft}p{1.2cm} >{\raggedleft}p{1.2cm} >{\raggedleft}p{1.2cm} >{\raggedleft}p{1.2cm} >{\raggedleft}p{1.2cm} >{\raggedleft}p{1.2cm} p{1.2cm}<{\raggedleft}}   \\[-0.5em] 	
			&\multicolumn{4}{c}{\textbf{\emph{Women with career in 2010}}}  &\multicolumn{4}{c}{\textbf{\emph{Women with no career in 2010}}}    \\
			&\multicolumn{1}{r}{{Mean}}&\multicolumn{1}{r}{{Std.Dev.}}&\multicolumn{1}{r}{{Min}}&\multicolumn{1}{r}{{Max}}&\multicolumn{1}{r}{{Mean}}&\multicolumn{1}{r}{{Std.Dev.}}&\multicolumn{1}{r}{{Min}}&\multicolumn{1}{r}{{Max}}\\
			\midrule
			age\_1       &        39.766&       3.054&          34&          44&      39.579&       3.098&          34&          44    \\   
			high educated\_1               &       0.607&       0.489&           0&           1&       0.097&       0.296&           0&           1\\    
			middle educated\_1 & 0.386 &    0.487 &   0  &  1 &  0.792 & 0.405 &  0  &  1  \\
			low educated\_1              &       0.005&       0.071&           0&           1&       0.097&       0.296&           0&           1\\ 
			non-German citizenship\_1         &       0.035&       0.185&           0&           1&       0.046&       0.210&           0&           1\\     
			career\_1            &       0.743&       0.437&           0&           1&       0.004&       0.062&           0&           1\\	
			employed\_1      &       0.982&       0.133&           0&           1&       0.868&       0.339&           0&           1   \\
			full-time employed\_1 &  0.919 & 0.272 &  0  &  1 &  0.390 & 0.485 & 0 &  1 \\
			part-time employed\_1   &       0.063&       0.242&           0&           1&       0.297&       0.457&           0&           1      \\	
			marginally employed\_1    &       0.000&       0.000&           0&           0&       0.181&       0.385&           0&           1     \\
			daily wage\_1 \newline (missing if non-employed)   & 163.283 &  27.507 &  13.041 &  177 &  50.080 &  39.624 &  0.029 & 177    \\
			daily wage\_1 \newline (zero if non-employed)   &     160.262&      35.032&           0&         177&      43.453&      40.624&           0&         177\\   
			experience\_1   &      15.619&       5.040&           0&          25&      16.387&       5.857&           0&          25  \\ 	\hline \\[-1.0em]
			\multicolumn{9}{l}{\small \textit{Source}: Own calculations based on data of the SIAB. }	\\
			\multicolumn{9}{l}{\small \textit{Notes}: $N_1=1\,567$ women with career in 2010 and $N_2=78\,215$ women with no career in 2010. $N =79\,782$}
		\end{tabular}
	\end{center}
\end{table*}
Table \ref{table3.1} provides the descriptive statistics of some key baseline variables. 
The table shows the descriptives separately for the  women with and without a professional career in 2010. 
The summary statistics show that the majority of women with no career in 2010 has vocational training (79\%), around 10\% are tertiary educated, and 10\% have no postsecondary degree. Unlike these, most of the women with a career in 2010 are highly educated (61\%), followed by women with middle education (37\%) and low education (0.5\%). 
Our sample of women has an average age of 40 years in 2009. 
The average employment rate is 87\% for women with no career in 2010 and 98\% for women with a career in 2010. 
Almost all women with a professional career in 2010 are full-time employed (92\%), followed by part-time working women (6\%). 
None of these women are marginally employed. On the other hand, only 39\% of women with no career in 2010 are full-time employed, followed by part-time (30\%) and marginally (18\%) employment. 
In our sample, around 4\% (5\%) of the women have non-German citizenship. 
The average daily wage among working women with a professional career in 2010 is around 163 EUR. 
However, for women without a professional career in 2010 it is only 50 EUR. 
Making a career is highly persistent, since 74\% of women with a professional career in 2010 also had a career in 2009. 
In the case of women without a career in 2010, however, it is only 0.4\%.

\section{Results}\label{Results}
How important is a woman's employment history for her professional career? 
To approach this question, we establish an FDR pipeline based on~\cite{Barber2015,Candes2016,Xie2019} to identify among a plethora of predictors derived from employment records those ones that are most predictive for a successful career.
 
In particular, we want to increase the estimation accuracy by effectively identifying the important predictor variables and improve the model interpretability.
For this purpose, we take into account a large set of variables derived from employment records, and apply a new aggregation scheme~\cite{Xie2019} for controlling the FDR to the data of the SIAB. 
We use this FDR approach beyond multiple testing for \textit{controlled} variable selection as introduced in~\cite{Barber2015} and~\cite{Candes2016}.
To achieve correct inference, we apply the knockoff-filter to the aggregation scheme which solves the controlled variable selection problem in such a generality and for a wide variety of test-statistics that it is attractive to introduce it in the field of economics.

We assume that our data generating process can be modeled precisely by using only a small, but a priori unknown number of predictor variables. 
Therefore we include all predictors to search for the truely important ones with the FDR pipeline.

\subsection{Main findings} 
Our main findings suggest that our data generating process can indeed be described by a sparse representation, since it improves the prediction performance and model interpretability. 
In particular, the (A)FDR methods show an improvement in terms of model interpretability and estimation accuracy compared to conventional variable selection methods such as the LASSO. 
In sum, we find that the employment status, working experience, daily wages, and wage increases in combination with a high level of education are truely associated with professional careers.
\subsection{Prediction and variable selection performance}
Our objective is to establish the FDR pipeline for variable selection in modern big data applications in economics. 
In doing so, we apply the proposed pipeline to labor market data and identify those variables from individual employment histories that are truely important for professional careers. 
Since there are no ground truths available for our application, that is, the true support $\mathcal{S}:=\text{supp}[\boldsymbol{\beta}]$ is unverifiable, we cannot measure variable selection accuracy directly. However, the ground truth is almost never available in empirical research. 
Therefore, we infer the methods' performance from the number of selected variables and the prediction accuracy.
We apply the different methods as described in the methodology section. 
In addition, we estimate conventional logistic regression with the full set of variables and without any variable (only intercept). 
These two latter estimations mark the two extreme situations in which all variables or no variable is relevant for predicting professional careers. 

For the ease of interpretation, we typically seek methods that yield a sparse model with a small number of variables. 
Moreover, we search for models with small prediction errors (good fit to the data). 
The model size is here crucial since our primary goal is variable selection. 
In particular, we are interested in the variable selection and prediction performances of the $l_1$-regularized logistic regression with 10-fold Cross-Validation refit. 
The prediction error of the 10-fold-CV with refit is defined as $\text{pred.~error}:= \frac{1}{10n^{\mathfrak{v}}}\sum_{j=1}^{10} \mathcal{L}[\boldsymbol{y}_{\mathcal{V}_j},X_{\mathcal{\hat{S}}_{\mathcal{V}_j}}\hat{\boldsymbol{\beta}}_{\text{refit}}[\boldsymbol{y}_{\mathcal{T}_j}, X_{\mathcal{\hat{S}}_{\mathcal{T}_j}}]]$, where $\mathcal{L}[\cdot]$ is the negative log-likelihood function, $X_{\mathcal{\hat{S}}}$ is the restricted design matrix after screening the non-zero coordinates with the proposed variable selection methods, $\hat{\boldsymbol{\beta}}_{\text{refit}}$ are the refitted estimators, and $\mathcal{T}_j:=\{1,\dots,n\} \backslash \mathcal{A}_j$ and $\mathcal{V}_j:= \mathcal{A}_j$, $j \in \{1, \dots, 10\}$ are our 10 training and validation sets, with $\mathcal{A}_1, \dots, \mathcal{A}_{10}$ having equal cardinality of $n/10$.
 
\begin{table*}[h!]
	\begin{center}
		\small
		\caption{\small Prediction performance. Our proposed method provides accurate prediction based on only 10 (19) predictors.}\label{table4.1}
		\begin{tabular}{p{4cm} >{\raggedleft}p{2cm}  p{3.5cm}<{\raggedleft} }   \\[-0.5em] 
		Method	&   Model size      & \multicolumn{1}{r}{Pred.~error  (10-fold-CV--refit)}  \\[0,2cm]
			LASSO & 42  & 1.47    \\[0,2cm]
			\textbf{FDR	$\text{LSM}$} &  \textbf{25} & \textbf{1.46} \\[0,2cm]
			\textbf{AFDR	$\text{LSM}$} &  \textbf{19}  &  \textbf{1.46} \\[0,2cm] 
			\textbf{FDR	$\text{LCD}_{\text{CV}}$} &  \textbf{2} & \textbf{1.42} \\[0,2cm] 
			\textbf{AFDR	$\text{LCD}_{\text{CV}}$} &  \textbf{10}  & \textbf{1.43} \\[0,2cm]    
			\textbf{FDR	$\text{LCD}_{\text{Testing}}$} &  \textbf{--}  & \textbf{--} \\[0,2cm] 
			\textbf{AFDR	$\text{LCD}_{\text{Testing}}$} &  \textbf{--} &  \textbf{--} \\[0,2cm]  
			Full &  327  & 1.47    \\[0,2cm] 
			Empty  &  0 & 7.71 \\
		\end{tabular}
	\end{center}
\end{table*}
Table \ref{table4.1} reports the model sizes and the prediction errors of the 10-fold-CV with refit.
First, it is noticable that the empty model (only intercept) has a large prediction error. 
This means that the employment and wage history variables are indeed able to improve the prediction quality and are correlated with the professional careers. 
On the other hand, the large-scale model with the full set of variables does not lead to a better fit of the data than the models selected by our proposed FDR pipeline. 
Thus, it is profitable to search for the main variables which improve the model fit and model interpretability.
Second, we observe that the LASSO method selects a considerably larger model than the (A)FDR control methods. 
This is expected, in view of LASSO-type estimators calibrated by cross-validation being designed for prediction and (A)FDR being designed for variable selection.
What is added to the aggregated knockoff filter over the usual LASSO are the data-driven test statistics to control the FDR. 
This additional feature typically results in a larger increase in accuracy after refitting. 
This can actually be observed in Table \ref{table4.2} that reports a slightly smaller refit error for the (A)FDR methods.
Third, by comparing the FDR methods with each other it seems that the (A)FDR control with CV is slightly superior to the other methods in terms of model size and prediction accuracy.
In contrast, (A)FDR control with the testing-based calibration is not usuable in our context since at least one variable should be selected.

Overall, no (A)FDR control method is significantly dominating in all measures, so that in empirical work the two aspects have to be weighted according to the primary goal. 
In our application, the model size is crucial since the primary objective is accurate variable selection, that is, the estimated support $\hat{\mathcal{S}}$ should be a good approximation of the true support $\mathcal{S}$. 
Below, we focus on the AFDR control methods for two main reasons. 
First, \cite{Xie2019} have shown in simulations that the aggregated knockoff filter can simultanously decrease the FDR and increase the power, while maintaining the original method's theoretical FDR guarantees. 
Second, we are interested in identifying a medium-sized model, because a model with only a few predictor variables leaves too little room for interpretation on the one hand, and on the other hand a model that is too large is difficult to interpret and harbors the risk of noise variables. 
Our results in Table \ref{table4.1} show that the AFDR method selects a medium-sized model regardless of the knockoff statistics chosen.

Next, we provide the variables selected by the different applied methods. 
Table \ref{table4.2} in the appendix contains the results for the six (A)FDR control estimations and Table \ref{table4.3} in the appendix displays the results for the LASSO estimation.
The upper panel of Table \ref{table4.2} displays the results for the (A)FDR control method based on the LSM statistic, whereas the two lower panels display the results based on the LCD statistic. 

We observe that variables that appear multiple times across methods are \texttt{wage profile 1}, \texttt{employed${}_i$}, \texttt{marginal employed$_3$}, \texttt{daily wage}${}_i$, \texttt{strong positive wage growth$_2$}, \texttt{change of establishment}${}_1$, \texttt{high educated}$_1$ $\times$ \texttt{change of establishment}${}_1$, \texttt{high educated}$_1$ $\times$ \texttt{part-time employed}$_j$, \texttt{high educated}$_1$ $\times$ \texttt{marginal employed}$_l$, \texttt{high educated}$_1$ $\times$ \texttt{daily wage}$_i$, \texttt{high educated}$_1$ $\times$ \texttt{strong positive wage growth}$_i$, with $i\in\{1,2,3,4,5\}$, $j\in\{1,5\}$, and $l\in\{2,3\}$. 
We can observe that the selected variables form two main groups: either predictors related to wages or predictors related to the employment status are selected. 
On the other hand, predictors related to unemployment benefits and welfare benefits are not selected by the methods at all and therefore are not associated with professional careers.

This finding is in line with our intuition: Daily wages and daily wages (or positive wage growths) combined with a high level of education are generally considered as important determinants for a professional career.
Our proposed FDR pipeline confirms this assumption, since all delays of the above mentioned variables are selected. The fact that several delays of the employment variable are selected has to do with the career prediction signal inherent in this variable. 
A continuous employment is almost indispensable for making a successful career. 
Finally, we observe that in the majority of cases where two baseline variables interact with each other, the high level of education interacts with the variables from the employment and wage history. 
This is also not suprisingly, given that a high educational level is often associated with a successful career. 
A closer look at these interactions reveals that the high level of education interacts with the variables that have already been selected as meaningful baseline predictors.

Tables \ref{table4.rob_1} and \ref{table4.rob_2} in the appendix contain the selected variables of the robustness checks. To check the robustness, we carried out the same analysis for the LSM knockoff statistics for the years 2009 and 2011. The results show that, although fewer variables are selected overall, the same variables as in the 2010 sample are selected. 

Overall, our results suggest that the proposed FDR pipeline is very useful in practice, especially if the researcher is faced with huge amounts of data. 
A distinctive feature of the FDR pipeline from conventional econometric tools is that it teases apart important from irrelevant variables while guaranteeing type I error control. Thus, the FDR approach enable economists to limit the number of covariates in a data-driven manner rather than by hand, in a way that the interpretability of economic models is preserved and correct inference is achieved. 
Therefore, we recommend economists who handle large amounts of data to use modern high-dimensional tools for variable selection. 
Due to the shift in the scale of economic data towards Big Data these high-dimensional models will become a new important toolbox in empirical economic reasearch alongside the conventional econometric toolbox. 
Having said this, a reasonable question is why should economists include so many covariates in the analysis? 
The answer is twofold. First, because we can due to the newly available large-scale administrative data sets.
Second, and more importantly, even though economists may believe that a certain economic outcome depends on a small fraction of all available variables, we have a priori no idea which ones are the truely important ones.

\subsection{Inference}
In the previous paragraph, we showed that the (A)FDR control, which was originally designed for multiple testing, is extremely useful for variable selection in empirical applications where a plethora of predictor variables is available.
The final goal of variable selection with (A)FDR control is to draw correct inferences about the estimates of the selected variables.
Therefore, in this paragraph we focus on the inference and interpretation of the estimates. Since the $l_1$ penalty sets a certain fraction of parameters exactly to zero and favors small estimates, this can lead to an overall unwanted shrinkage of the estimates.
To remove such biases, we refit the penalized estimators with subsequent logistic regression and least-squares estimation on the support. 
The advantage of this two-step procedure is that least-squares provides an accurate and unbiased estimator in low-dimensional and correctly specified models which can be interpreted straightforward. 
In addition, we calculate marginal effects of the logistic regression estimates to compare them with the respective least-squares estimates.

In Table \ref{table4.4} we provide the estimation results for refitting AFDR $\text{LCD}_{\text{CV}}$. 
Tables \ref{table4.5}, \ref{Tab: 3} and \ref{Tab: 4} in the appendix report the corresponding estimation results for refitting AFDR LSM, FDR $\text{LCD}_{\text{CV}}$ and FDR LSM. 
The respective tables provide the OLS estimates (with refit), the logistic regression estimates (with refit) and the corresponding marginal effects. 

A first look at the results shows that the estimates are robust against the respective specification, and that most estimates are statistically significant. 
The OLS and logistic regression results in columns 3 and 5 are very similar. 
For the sake of simplicity, we only interpret the statistically significant OLS estimates in the second column of Table \ref{table4.4}, since only the continuous variables are standardized in this column, so that a reasonable interpretation of the binary variables is possible. 
In addition, we only give an exemplary interpretation of one lagged term from the respective variable group.
\begin{table*}[hp]
	\begin{center}
		\small
		\caption{\small Inference}\label{table4.4}
		\begin{tabular}{lSSSS}   \\[-0.5em] 
			$\text{LCD}_{CV}$ &    \\[0,2cm] 
			$\boldsymbol{\hat{\mathcal{S}}}_{\text{AFDR}}$ & \text{OLS Estimate}$^{++}$   & \text{OLS Estimate}$^+$  & \text{Estimate}$^+$  & \text{Marginal effect}$^+$   \\
			& \text{(with refit)} & \text{(with refit)} & \text{(with refit)} &\text{(with refit)} \\[0,3cm]
			\text{daily wage}${}_1$   & 0.0404***  & 0.0404***    & 0.1647***  & 0.0406***   \\
			& 0.0012 & 0.0012 & 0.0191 & 0.0047 \\
			\text{daily wage}${}_2$ & 0.0081***  &  0.0081***  & 0.0321  & 0.0079   \\
			& 0.0013 & 0.0013 & 0.0206 &  0.0051 \\
			\text{daily wage}${}_4$ & 0.0146***   & 0.0146***  & 0.0600***  & 0.0148*** \\
			& 0.0012 &  0.0012 & 0.0188 & 0.0046 \\
			\text{daily wage}${}_5$ & 0.0064*** &  0.0064*** & 0.0258*  & 0.0064*   \\
			& 0.0010 & 0.0010 & 0.0153 & 0.0038 \\
			\text{employed}${}_1$ & -0.0031  & -0.0010  & -0.0042 & -0.0010  \\
			&  0.0022 & 0.0008 &  0.0120 & 0.0030  \\
			\text{employed}${}_2$ &  -0.0158***  & -0.0055***  & -0.0222**  & -0.0055**  \\
			& 0.0018 & 0.0006 & 0.0100 & 0.0025  \\
			\text{employed}${}_4$ &  -0.0306*** & -0.0130***  & -0.0529***  & -0.0130***  \\
			& 0.0014 & 0.0006 & 0.0094 & 0.0023  \\
			\text{strong negative wage growth}${}_1$ & 0.0136***   &  0.0060***  & 0.0245***  & 0.0060***   \\
			& 0.0013 & 0.0006 & 0.0089 & 0.0022 \\
			\text{strong negative wage growth}${}_2$ & 0.0120***  &  0.0049***  & 0.0199** & 0.0049**  \\ 
			& 0.0012 & 0.0005 & 0.0078  & 0.0019 \\
			\text{experience}${}_1$ $\times$ \text{employed}${}_1$ &  -0.0012***  & -0.0094***   & -0.0385***  & -0.0095***  \\
			& 0.0001 & 0.0007 & 0.0109 & 0.0027   \\[0,2cm]
			\multicolumn{5}{l}{$^+$ Variables are standardized.} \\
			\multicolumn{5}{l}{$^{++}$ Only continuous baseline variables are standardized.} \\
			\multicolumn{5}{l}{$^*$, $^{**}$, $^{***}$ statistically significant at the 10\%, 5\% and 1\% levels, respectively.}
				\end{tabular}
		\end{center}
	\end{table*}  
			
Starting to interpret the OLS estimate on $\texttt{daily wage}_1$, we find that, by holding all other factors constant (h.a.o.f.c.),  increasing the daily wage in 2009 by one standard deviation leads on average to a 4.0 standard deviation higher probability of making a professional career in 2010. 
It is striking that all lagged terms of the daily wage have a significant positive impact on the professional career.
Continuing with the next group of selected variables that has statistically significant estimates, we find that being employed in 2008 decreases the probability of making a professional career in 2010 on average by 1.6 percentage points compared of being non-employed in 2008 (h.a.o.f.c.). 
Moreover, we find that having a strong negative wage growth in 2009 increases the probability of making a successful career in 2010 on average by 1.4 percentage points compared of having a normal or a positive wage growth in 2009. 
Finally, we give the interpretation on the interaction term $\texttt{experience}_1 \times \texttt{employed}_1$: h.a.o.f.c., women who are employed in 2009 have on average a 0.1 standard deviation lower probability of making a professional career in 2010 than women who are unemployed in 2009 with every additional standard deviation experience.

Table \ref{table4.5} in the appendix gives further important insights regarding the interpretation of the OLS estimates of $\mathcal{\hat{S}}_{\text{AFDR}}$ with the $\text{LSM}$ statistic. 
We find that being employed in 2009 increases the probability of making a professional career in 2010 on average by 1.2 percentage points compared of being non-employed in 2009 (h.a.o.f.c). 
We further find that being marginal employed in 2007 decreases the probability of making a professional career in 2010 on average by 0.86 percentage point compared of being full-time, part-time or unemployed in 2007. 
On the other hand, a steady wage growth over the past five years increases the probability of making a professional career on average by 0.55 percentage point (h.a.o.f.c.).


\section{Conclusion}
Based on an empirical application towards the labor market, this paper introduced the potentials of high-dimensional tools aimed to \textit{controlled} variable selection to economists. More specifically, we applied a new aggregation scheme for FDR control \cite{Xie2019} to a high-dimensional logistic regression model, which teases apart important from irrelevant variables while maximizing the (statistical) power and guaranteeing Type I error control. So far, the aggreagated FDR control method has only been used in the context of high-dimensional linear regression models \cite{Xie2019}. However, we were easily able to extent the aggregation scheme for the non-linear case by exploiting the framework of MX knockoffs \cite{Candes2016}. The MX knockoff-filter effectively controls the aggregated false discovery rate and performs valid inference in our high-dimensional logistic regression model by mimicking the correlation structure found within the potential covariates.

In particular, we applied the MX knockoff-filter to the data from the Sample of Integrated Employment Biographies to discover which variables from individual employment histories are truely associated with female professional careers. We assumed that the binary career outcome variable can be modeled accurately by using a \textit{sparse} representation of the covariates, but unlike the conventional economic literature, we were unsure about the identity of the relevant covariates and selected them in a data-driven manner. To reach a parsimonious model, we used the $l_1$ penalized technique to solve the logistic regression throughout our estimations.

Our main results suggest that our high-resolution logistic regression model can indeed be described by a sparse representation, since it improves the prediction performance and model interpretability. Indeed, the (A)FDR methods, which are designed for variable selection, show an improvement in terms of estimation accuracy and model interpretability compared to conventional variable selection methods such as the LASSO, which are designed for prediction. Overall, the relatively small subset of the employment history variables that are genuinely related to female professional careers includes the working experience, employment status, daily wage and wage increases in combination with a high level of education.

Our results provide new insights for the empirical work in the economic discipline. First, the high-dimensional tools presented here enable economists to fit high-dimensional models. Due to the newly available large-scale data-sets these high-dimensional models will become a new important workhorse in empirical economic research along the conventional econometrics toolbox. Second, based on the empirical application towards the labor market, we have shown that data-driven (A)FDR control, which was originally designed for multiple testing, is also useful for variable selection and correct inference in high-dimensional settings. This is particularly an important insight given the fact that there is less knowledge about variable selection and certainly not about (A)FDR control  in the economic literature. Third, the tools for \textit{controlled} variable selection introduced here enable economists to limit the number of explanatory variables in a data-driven fashion, in a way that the interpretability of economic models is preserved by searching for a sparse representation of the data generating process.

\bibliography{Literatur} 
\bibliographystyle{alphadin}
\renewcommand\refname{Literaturverzeichnis} 

\section{Appendix}
\singlespacing
\begin{longtable}[!htbp]{p{5cm}  >{\raggedleft}p{1.2cm} >{\raggedleft}p{1.2cm}  >{\raggedleft}p{1cm} >{\raggedleft}p{1cm} >{\raggedleft}p{1.2cm} >{\raggedleft}p{1.2cm}  >{\raggedleft}p{1cm} p{1cm}<{\raggedleft}}  \hline \\[-0.5em] 
	\caption{ Descriptives of Baseline Predictors}\label{Tab:1}\\[0,5cm] \toprule \\[-0.5em] 
	\endfirsthead	
	\caption*{\textbf{Table \ref{Tab:1}:} Continued}\\\toprule
	\endhead
	\endfoot
	\bottomrule
	\endlastfoot
	&\multicolumn{4}{c}{\textbf{\emph{Women with career in 2010}}}   &\multicolumn{4}{c}{\textbf{\emph{Women with no career in 2010}}}\\
	&\multicolumn{1}{c}{{Mean}}&\multicolumn{1}{c}{{Std.Dev.}}&\multicolumn{1}{l}{{Min}}&\multicolumn{1}{l}{{Max}}&\multicolumn{1}{c}{{Mean}}&\multicolumn{1}{c}{{Std.Dev.}}&\multicolumn{1}{l}{{Min}}&\multicolumn{1}{l}{{Max}}\\[0,1cm]
	\midrule \\[-0.5em] 
career              &       1.000&       0.000&           1&           1&       0.000&       0.000&           0&           0\\
age$_1$               &      39.766&       3.054&          34&          44&      39.579&       3.098&          34&          44\\
age$^2_1$              &    1590.641&     239.804&        1156&        1936&    1576.066&     242.984&        1156&        1936\\
high educated$_1$              &       0.607&       0.489&           0&           1&       0.097&       0.296&           0&           1\\
low educated$_1$               &       0.005&       0.071&           0&           1&       0.097&       0.296&           0&           1\\
non-German citizenship$_1$           &       0.035&       0.185&           0&           1&       0.046&       0.210&           0&           1\\
employed$_1$            &       0.982&       0.133&           0&           1&       0.868&       0.339&           0&           1\\
employed$_2$            &       0.972&       0.165&           0&           1&       0.858&       0.349&           0&           1\\
employed$_3$            &       0.962&       0.192&           0&           1&       0.808&       0.394&           0&           1\\
employed$_4$            &       0.944&       0.229&           0&           1&       0.762&       0.426&           0&           1\\
employed$_5$            &       0.941&       0.236&           0&           1&       0.727&       0.445&           0&           1\\
part-time employed$_1$          &       0.063&       0.242&           0&           1&       0.297&       0.457&           0&           1\\
part-time employed$_2$          &       0.064&       0.245&           0&           1&       0.276&       0.447&           0&           1\\
part-time employed$_3$          &       0.062&       0.241&           0&           1&       0.248&       0.432&           0&           1\\
part-time employed$_4$          &       0.066&       0.248&           0&           1&       0.223&       0.416&           0&           1\\
part-time employed$_5$          &       0.059&       0.236&           0&           1&       0.203&       0.402&           0&           1\\
marginally employed$_1$           &       0.000&       0.000&           0&           0&       0.181&       0.385&           0&           1\\
marginally employed$_2$           &       0.003&       0.050&           0&           1&       0.186&       0.389&           0&           1\\
marginally employed$_3$           &       0.003&       0.050&           0&           1&       0.179&       0.383&           0&           1\\
marginally employed$_4$           &       0.003&       0.056&           0&           1&       0.168&       0.374&           0&           1\\
marginally employed$_5$           &       0.003&       0.050&           0&           1&       0.159&       0.366&           0&           1\\
daily wage$_1$              &     160.262&      35.032&           0&         177&      43.453&      40.624&           0&         177\\
daily wage$_2$              &     152.078&      39.031&           0&         173&      42.213&      40.222&           0&         173\\
daily wage$_3$              &     143.501&      43.241&           0&         172&      38.830&      39.023&           0&         172\\
daily wage$_4$              &     139.335&      47.463&           0&         172&      36.855&      39.148&           0&         172\\
daily wage$_5$              &     132.962&      48.308&           0&         170&      35.325&      38.724&           0&         170\\
wage profile 1           &       0.143&       0.350&           0&           1&       0.068&       0.253&           0&           1\\
wage profile 2           &       0.726&       0.446&           0&           1&       0.679&       0.467&           0&           1\\
years since last change of establishment$_1$&       4.803&       4.689&           0&          22&       4.560&       5.001&           0&          23\\
years since last change of establishment$_2$&       4.446&       4.446&           0&          21&       4.217&       4.769&           0&          22\\
years since last change of establishment$_3$&       4.168&       4.185&           0&          21&       4.064&       4.546&           0&          21\\
years since last change of establishment$_4$&       3.832&       3.917&           0&          20&       3.861&       4.325&           0&          20\\
years since last change of establishment$_5$&       3.474&       3.647&           0&          19&       3.588&       4.106&           0&          19\\
change of establishment$_1$   &       0.122&       0.327&           0&           1&       0.148&       0.355&           0&           1\\
change of establishment$_2$   &       0.143&       0.350&           0&           1&       0.178&       0.383&           0&           1\\
change of establishment$_3$   &       0.135&       0.342&           0&           1&       0.165&       0.371&           0&           1\\
change of establishment$_4$   &       0.137&       0.344&           0&           1&       0.151&       0.358&           0&           1\\
change of establishment$_5$   &       0.149&       0.356&           0&           1&       0.148&       0.355&           0&           1\\
strong negative wage \newline growth$_1$&       0.174&       0.379&           0&           1&       0.266&       0.442&           0&           1\\
strong negative wage \newline growth$_2$ &       0.172&       0.378&           0&           1&       0.213&       0.409&           0&           1\\
strong negative  wage \newline growth$_3$&       0.209&       0.406&           0&           1&       0.216&       0.412&           0&           1\\
strong negative wage \newline growth$_4$ &       0.181&       0.385&           0&           1&       0.214&       0.410&           0&           1\\
strong negative wage \newline growth$_5$ &       0.195&       0.397&           0&           1&       0.222&       0.416&           0&           1\\
strong positive wage \newline growth$_1$ &       0.239&       0.427&           0&           1&       0.320&       0.467&           0&           1\\
strong positive wage \newline growth$_2$ &       0.293&       0.455&           0&           1&       0.337&       0.473&           0&           1\\
strong positive wage \newline growth$_3$ &       0.285&       0.451&           0&           1&       0.290&       0.454&           0&           1\\
strong positive wage \newline growth$_4$ &       0.313&       0.464&           0&           1&       0.254&       0.435&           0&           1\\
strong positive wage \newline growth$_5$ &       0.334&       0.472&           0&           1&       0.235&       0.424&           0&           1\\
experience$_1$             &      15.619&       5.040&           0&          25&      16.387&       5.857&           0&          25\\
experience$_2$             &      14.620&       5.038&           0&          24&      15.401&       5.818&           0&          24\\
experience$_3$             &      13.622&       5.031&           0&          23&      14.422&       5.765&           0&          23\\
experience$_4$             &      12.630&       5.011&           0&          22&      13.453&       5.689&           0&          22\\
experience$_5$             &      11.641&       4.984&           0&          21&      12.494&       5.596&           0&          21\\
experience$^2_1$            &     269.337&     155.428&           0&         625&     302.842&     170.909&           0&         625\\
experience$^2_2$            &     239.098&     145.550&           0&         576&     271.053&     159.602&           0&         576\\
experience$^2_3$          &     210.856&     135.706&           0&         529&     241.229&     148.408&           0&         529\\
experience$^2_4$           &     184.604&     125.912&           0&         484&     213.354&     137.358&           0&         484\\
experience$^2_5$           &     160.334&     116.186&           0&         441&     187.408&     126.485&           0&         441\\
days of unemployment \newline benefits$_1$         &       1.378&      14.112&           0&         350&      15.678&      55.186&           0&         365\\
days of unemployment \newline benefits$_2$        &       1.068&      12.133&           0&         291&      14.863&      53.283&           0&         366\\
days of unemployment \newline benefits$_3$         &       1.588&      14.779&           0&         273&      20.761&      69.120&           0&         365\\
days of unemployment \newline benefits$_4$        &       2.149&      20.287&           0&         363&       5.104&      29.957&           0&         365\\
days of unemployment \newline benefits$_5$         &       2.383&      17.683&           0&         244&       7.837&      37.985&           0&         365\\
days of registered job search \newline while not employed$_1$       &       0.114&       2.884&           0&          97&       4.838&      28.582&           0&         365\\
days of registered job search \newline while not employed$_2$       &       0.208&       3.702&           0&         105&       5.010&      29.265&           0&         366\\
days of registered job search \newline while not employed$_3$     &       0.035&       0.870&           0&          31&       6.011&      34.973&           0&         365\\
days of registered job search \newline while not employed$_4$     &       0.000&       0.000&           0&           0&       0.000&       0.000&           0&           0\\
days of registered job search \newline while not employed$_5$      &       0.000&       0.000&           0&           0&       0.000&       0.000&           0&           0\\
unemployment benefits$_1$               &       0.001&       0.036&           0&           1&       0.037&       0.189&           0&           1\\
unemployment benefits$_2$                &       0.001&       0.036&           0&           1&       0.033&       0.179&           0&           1\\
unemployment benefits$_3$                 &       0.002&       0.044&           0&           1&       0.051&       0.220&           0&           1\\
unemployment benefits$_4$                &       0.004&       0.067&           0&           1&       0.010&       0.099&           0&           1\\
unemployment benefits$_5$               &       0.003&       0.050&           0&           1&       0.016&       0.127&           0&           1\\
registered job search \newline while not employed$_1$            &       0.000&       0.000&           0&           0&       0.009&       0.094&           0&           1\\
registered job search \newline while not employed$_2$              &       0.000&       0.000&           0&           0&       0.009&       0.095&           0&           1\\
registered job search \newline while not employed$_3$             &       0.000&       0.000&           0&           0&       0.012&       0.109&           0&           1\\
registered job search \newline while not employed$_4$             &       0.000&       0.000&           0&           0&       0.000&       0.000&           0&           0\\
registered job search \newline while not employed$_5$            &       0.000&       0.000&           0&           0&       0.000&       0.000&           0&           0\\
occupation1\_1      &       0.000&       0.000&           0&           0&       0.010&       0.098&           0&           1\\
occupation2\_1      &       0.000&       0.000&           0&           0&       0.000&       0.022&           0&           1\\
occupation3\_1      &       0.000&       0.000&           0&           0&       0.001&       0.031&           0&           1\\
occupation4\_1      &       0.003&       0.050&           0&           1&       0.006&       0.077&           0&           1\\
occupation5\_1      &       0.000&       0.000&           0&           0&       0.004&       0.066&           0&           1\\
occupation6\_1      &       0.000&       0.000&           0&           0&       0.000&       0.021&           0&           1\\
occupation7\_1      &       0.000&       0.000&           0&           0&       0.003&       0.052&           0&           1\\
occupation8\_1      &       0.000&       0.000&           0&           0&       0.006&       0.076&           0&           1\\
occupation9\_1      &       0.001&       0.025&           0&           1&       0.007&       0.081&           0&           1\\
occupation10\_1     &       0.000&       0.000&           0&           0&       0.007&       0.083&           0&           1\\
occupation11\_1     &       0.000&       0.000&           0&           0&       0.004&       0.066&           0&           1\\
occupation12\_1     &       0.001&       0.036&           0&           1&       0.025&       0.156&           0&           1\\
occupation13\_1     &       0.000&       0.000&           0&           0&       0.001&       0.023&           0&           1\\
occupation14\_1     &       0.000&       0.000&           0&           0&       0.001&       0.028&           0&           1\\
occupation15\_1     &       0.000&       0.000&           0&           0&       0.000&       0.021&           0&           1\\
occupation16\_1     &       0.000&       0.000&           0&           0&       0.001&       0.029&           0&           1\\
occupation17\_1     &       0.001&       0.025&           0&           1&       0.015&       0.120&           0&           1\\
occupation18\_1     &       0.000&       0.000&           0&           0&       0.014&       0.119&           0&           1\\
occupation19\_1     &       0.000&       0.000&           0&           0&       0.000&       0.019&           0&           1\\
occupation20\_1     &       0.064&       0.246&           0&           1&       0.007&       0.083&           0&           1\\
occupation21\_1     &       0.032&       0.176&           0&           1&       0.016&       0.127&           0&           1\\
occupation22\_1     &       0.082&       0.275&           0&           1&       0.116&       0.320&           0&           1\\
occupation23\_1     &       0.108&       0.310&           0&           1&       0.030&       0.172&           0&           1\\
occupation24\_1     &       0.029&       0.169&           0&           1&       0.045&       0.206&           0&           1\\
occupation25\_1     &       0.477&       0.500&           0&           1&       0.269&       0.444&           0&           1\\
occupation26\_1     &       0.020&       0.141&           0&           1&       0.010&       0.099&           0&           1\\
occupation27\_1     &       0.033&       0.178&           0&           1&       0.010&       0.101&           0&           1\\
occupation28\_1     &       0.108&       0.311&           0&           1&       0.113&       0.317&           0&           1\\
occupation29\_1     &       0.031&       0.172&           0&           1&       0.083&       0.275&           0&           1\\
occupation30\_1     &       0.003&       0.056&           0&           1&       0.139&       0.346&           0&           1\\  \hline \\[-1.0em]
	\multicolumn{9}{l}{\small \textit{Notes}: $N_1=1\,567$ career women and $N_2=78\,215$ non-career women in 2010.}
\end{longtable}

\begin{table*}
	\begin{center}
		\small
		\caption{\small Selected variables with (A)FDR control}\label{table4.2}
		\begin{tabular}[!htbp]{p{8cm} p{8cm} } \hline  \\[-0.5em] 
			\multicolumn{2}{c}{LSM statistic}   \\    \cline{1-2} \\[-0.5em] 
			$\boldsymbol{\hat{\mathcal{S}}}_{\text{FDR}}$  & $\boldsymbol{\hat{\mathcal{S}}}_{\text{AFDR}}$      \\[0,2cm] 
			$\text{employed}_1$  & $\text{employed}_1$   \\
			$\text{marginal employed}_3$ & $\text{marginal employed}_3$   \\
			$\text{change of establishment}_1$ & $\text{change of establishment}_1$   \\
			$\text{change of establishment}_4$  & --   \\
			$\text{change of establishment}_5$ & --   \\
			$\text{strong negative wage growth}_5$  & --  \\
			$\text{strong positive wage growth}_2$ & $\text{strong positive wage growth}_2$    \\
			$\text{wage profile 1}$	& $\text{wage profile 1}$   \\
			$\text{high educated}_1 \times \text{part-time employed}_1$& $\text{high educated}_1 \times \text{part-time employed}_1$                  \\
			$\text{high educated}_1 \times \text{part-time employed}_5$ & $\text{high educated}_1 \times \text{part-time employed}_5$                  \\
			$\text{high educated}_1 \times \text{marginal employed}_1$  &  --                \\
			$\text{high educated}_1 \times \text{marginal employed}_2$ & $\text{high educated}_1 \times \text{marginal employed}_2$                  \\
			$\text{high educated}_1 \times \text{marginal employed}_3$  &    --              \\
			$\text{high educated}_1 \times \text{strong positive wage growth}_1$& $\text{high educated}_1 \times \text{strong positive wage growth}_1$        	\\ 
			$\text{high educated}_1 \times \text{strong positive wage growth}_2$     & $\text{high educated}_1 \times \text{strong positive wage growth}_2$        	\\ 
			$\text{high educated}_1 \times \text{strong positive wage growth}_3$    & $\text{high educated}_1 \times \text{strong positive wage growth}_3$        	\\ 
			$\text{high educated}_1 \times \text{strong positive wage growth}_4$     & $\text{high educated}_1 \times \text{strong positive wage growth}_4$        	\\ 
			$\text{high educated}_1 \times \text{strong positive wage growth}_5$   & $\text{high educated}_1 \times \text{strong positive wage growth}_5$        	\\
			--  & $\text{high educated}_1 \times \text{employed}_1$       	\\ 
			$\text{high educated}_1 \times \text{change of establishment}_1$&  $\text{high educated}_1 \times \text{change of establishment}_1$       	\\                                                 
			$\text{high educated}_1 \times \text{daily wage}_1$& $\text{high educated}_1 \times \text{daily wage}_1$        	\\ 
			$\text{high educated}_1 \times \text{daily wage}_3$&  $\text{high educated}_1 \times \text{daily wage}_3$       	\\ 
			$\text{high educated}_1 \times \text{daily wage}_4$        &  $\text{high educated}_1 \times \text{daily wage}_4$       	\\
			$\text{high educated}_1 \times \text{daily wage}_5$          &  $\text{high educated}_1 \times \text{daily wage}_5$       	\\     
			$\text{experience}_2 \times \text{strong positive wage growth}_2$ &   --         \\
			$\text{experience}_5 \times \text{daily wage}_5$   &  --        \\   \hline \\[-0.5em]
			\multicolumn{2}{c}{$\text{LCD}_{CV}$ statistic}                        \\ \cline{1-2}  \\[-0.5em] 
			$\boldsymbol{\hat{\mathcal{S}}}_{\text{FDR}}$  & $\boldsymbol{\hat{\mathcal{S}}}_{\text{AFDR}}$ \\ 
			$\text{daily wage}_1$  &   $\text{daily wage}_1$ \\
			$\text{daily wage}_2$  &   $\text{daily wage}_2$ \\
			& $\text{daily wage}_4$ \\
			--  &   $\text{daily wage}_5$ \\
			-- & $\text{employed}_1$   \\
			-- & $\text{employed}_2$   \\
			-- & $\text{employed}_4$   \\	
			--  & $\text{strong negative wage growth}_1$   \\ 	
			--  & $\text{strong negative wage growth}_2$   \\ 
			&  $\text{experience}_1 \times \text{employed}_1$   \\ \hline \\[-0.5em]
			\multicolumn{2}{c}{$\text{LCD}_{AV}$ statistic} \\ \cline{1-2}  \\[-0.5em]
			$\boldsymbol{\hat{\mathcal{S}}}_{\text{FDR}}$ &  $\boldsymbol{\hat{\mathcal{S}}}_{\text{AFDR}}$  \\ 
			--& --  \\[0,2cm] \hline \\[-0.5em]
			\multicolumn{2}{l}{\textit{Notes:} Target level $q=0.10$. } 		
		\end{tabular}
	\end{center}
\end{table*}

\begin{table*}
	\begin{center}
		\small
		\caption{\small Robustness check 1: Selected variables with (A)FDR control}\label{table4.rob_1}
		\begin{tabular}[!htbp]{p{8cm} p{8cm} } \hline  \\[-0.5em] 
			\multicolumn{2}{c}{LSM statistic}   \\    \cline{1-2} \\[-0.5em] 
			$\boldsymbol{\hat{\mathcal{S}}}_{\text{FDR}}$  & $\boldsymbol{\hat{\mathcal{S}}}_{\text{AFDR}}$      \\[0,2cm] 
        $\text{employed}_3$  & -- \\
        $\text{change of establishment}_1$ & -- \\
        $\text{strong positive wage growth}_1$  & -- \\
        $\text{high educated}_1 \times \text{part-time employed}_1$ & $\text{high educated}_1 \times \text{part-time employed}_1$ \\
        $\text{high educated}_1 \times \text{marginal employed}_2$ & --  \\
        $\text{high educated}_1 \times \text{marginal employed}_4$ & --  \\
        $\text{high educated}_1 \times \text{strong negative wage growth}_1$ & -- \\
        $\text{high educated}_1 \times \text{strong positive wage growth}_1$ &  $\text{high educated}_1 \times \text{strong positive wage growth}_1$ \\
        $\text{high educated}_1 \times \text{strong positive wage growth}_2$ & -- \\
        $\text{high educated}_1 \times \text{strong positive wage growth}_3$ & -- \\
        $\text{high educated}_1 \times \text{strong positive wage growth}_4$ & -- \\
        $\text{high educated}_1 \times \text{wage profile 1}$ & $\text{high educated}_1 \times \text{wage profile 1}$  \\
        $\text{experience}_4 \times \text{daily wage}_4$ & -- \\ \\ \hline \\[-0.5em]
        \multicolumn{2}{l}{\textit{Notes:} Target level $q=0.10$. 2009 Sample } 		
		\end{tabular}
	\end{center}
\end{table*}
\begin{table*}
	\begin{center}
		\small
		\caption{\small Robustness check 2: Selected variables with (A)FDR control}\label{table4.rob_2}
		\begin{tabular}[!htbp]{p{8cm} p{8cm} } \hline  \\[-0.5em] 
			\multicolumn{2}{c}{LSM statistic}   \\    \cline{1-2} \\[-0.5em] 
			$\boldsymbol{\hat{\mathcal{S}}}_{\text{FDR}}$  & $\boldsymbol{\hat{\mathcal{S}}}_{\text{AFDR}}$      \\[0,2cm] 
			$\text{high educated}_1 \times \text{part-time employed}_1$ & $\text{high educated}_1 \times \text{part-time employed}_1$ \\			
			$\text{high educated}_1 \times \text{part-time employed}_5$	& $\text{high educated}_1 \times \text{part-time employed}_5$ \\
			$\text{high educated}_1 \times \text{strong positive wage growth}_1$ & $\text{high educated}_1 \times \text{strong positive wage growth}_1$ \\
			$\text{high educated}_1 \times \text{strong positive wage growth}_2$ & $\text{high educated}_1 \times \text{strong positive wage growth}_2$\\
			$\text{high educated}_1 \times \text{strong positive wage growth}_3$ & $\text{high educated}_1 \times \text{strong positive wage growth}_3$\\
			$\text{high educated}_1 \times \text{strong positive wage growth}_4$ & $\text{high educated}_1 \times \text{strong positive wage growth}_4$ \\
			$\text{high educated}_1 \times \text{strong positive wage growth}_5$ & $\text{high educated}_1 \times \text{strong positive wage growth}_5$\\
			-- & $\text{high educated}_1 \times \text{change of establishment}_1$ \\
			$\text{high educated}_1 \times \text{daily wage}_1$ & $\text{high educated}_1 \times \text{daily wage}_1$ \\
			$\text{high educated}_1 \times \text{daily wage}_3$	& $\text{high educated}_1 \times \text{daily wage}_3$ \\
			$\text{high educated}_1 \times \text{daily wage}_5$	& 	$\text{high educated}_1 \times \text{daily wage}_5$ \\
				\\ \hline \\[-0.5em]

			\multicolumn{2}{l}{\textit{Notes:} Target level $q=0.10$. 2011 Sample } 		
		\end{tabular}
	\end{center}
\end{table*}
\begin{table*}
	\begin{center}
		\small
		\caption{\small Selected variables with LASSO}\label{table4.3}
		\begin{tabular}[!htbp]{p{8cm} p{8cm}} \hline  \\[-0.5em] 
			No statistic    \\    \\[-0.5em] 
			$\boldsymbol{\hat{\mathcal{S}}}_{\text{LASSO}}$      \\[0,2cm] 
			$\text{daily wage}\_{1}$ & $\text{high educated}_1 \times \text{daily wage}\_{1}$\\ 
			$\text{daily wage}\_{2}$ & $\text{high educated}_1 \times \text{daily wage}\_{2}$\\
			$\text{daily wage}\_{3}$ & $\text{high educated}_1 \times \text{daily wage}\_{3}$ \\
			$\text{daily wage}\_{4}$ & $\text{high educated}_1 \times \text{daily wage}\_{4}$\\
			$\text{daily wage}\_{5}$ & $\text{high educated}_1 \times \text{daily wage}\_{5}$ \\
			$\text{employed}\_{1}$ & $\text{experience}_3 \times \text{daily wage}\_{3}$\\
			$\text{employed}\_{3}$ & $\text{experience}_5 \times \text{daily wage}\_{5}$ \\
			$\text{employed}\_{4}$ \\
			$\text{employed}\_{5}$ \\
			$\text{marginally employed}\_{1}$ \\
			$\text{marginally employed}\_{2}$ \\
			$\text{marginally employed}\_{3}$ \\
			$\text{marginally employed}\_{4}$ \\
			$\text{marginally employed}\_{5}$ \\
			$\text{years since last change of establishment}\_{1}$ \\
			$\text{change of establishment}\_{1}$ \\
			$\text{change of establishment}\_{2}$ \\
			$\text{strong positive wage growth}\_{1}$ \\
			$\text{strong positive wage growth}\_{2}$ \\
			$\text{occupation29}\_{1}$ \\
			$\text{wage profile 1}$ \\
			$\text{high educated}_1 \times \text{part-time employed}\_{1}$ \\
			$\text{high educated}_1 \times \text{part-time employed}\_{5}$ \\
			$\text{high educated}_1 \times \text{marginally employed}\_{2}$ \\
			$\text{high educated}_1 \times \text{marginally employed}\_{3}$ \\
			$\text{high educated}_1 \times \text{strong negative wage growth}\_{1}$ \\
			$\text{high educated}_1 \times \text{strong positive wage growth}\_{1}$ \\
			$\text{high educated}_1 \times \text{strong positive wage growth}\_{2}$ \\
			$\text{high educated}_1 \times \text{strong positive wage growth}\_{3}$ \\
			$\text{high educated}_1 \times \text{strong positive wage growth}\_{4}$ \\
			$\text{high educated}_1 \times \text{strong positive wage growth}\_{5}$ \\
			$\text{high educated}_1 \times \text{employed}\_{1}$ \\
			$\text{high educated}_1 \times \text{employed}\_{4}$ \\
			$\text{high educated}_1 \times \text{employed}\_{5}$ \\	
			$\text{high educated}_1 \times \text{change of establishment}\_{1}$ \\
			\\ \hline \\[-0.5em]			
		\end{tabular}
	\end{center}
\end{table*}
%

\begin{table*}
	\begin{center}
		\small
		\caption{\small Inference}\label{table4.5}
		\begin{tabular}[!htbp]{lSSSS}  \\[-0.5em] 
			$\text{LSM}$ &    \\[0,2cm]   
			$\boldsymbol{\hat{\mathcal{S}}}_{\text{AFDR}}$ &  \text{OLS Estimate}$^{++}$  & \text{OLS Estimate}$^+$   & \text{Estimate}$^+$  & \text{Marginal effect}$^+$  \\
			& \text{(with refit)} & \text{(with refit)} & \text{(with refit)} & \text{(with refit)} \\
			\text{employed}${}_1$  & 0.0121***   & 0.0035***  & 0.0142*  & 0.0035*   \\
			& 0.0013 & 0.0005 & 0.0074 & 0.0018 \\
			\text{marginal employed}${}_3$ & -0.0086***  & -0.0033*** & -0.0125*  & -0.0031*   \\
			& 0.0012 & 0.0004 & 0.0073 & 0.0018 \\
			\text{change of establishment}${}_1$ &  -0.0005  & 0.0001  & 0.0000  & 0.0000    \\
			&  0.0013 & 0.0004 & 0.0073 & 0.0018 \\
			\text{strong positive wage growth}${}_2$ & -0.0018*  & -0.0009* & -0.0036 & -0.0009     \\
			& 0.0010 &  0.0005 &  0.0077 & 0.0019 \\
			\text{wage profile 1}	& 0.0055***  & 0.0016***  & 0.0066 & 0.0016  \\
			& 0.0017 & 0.0004 & 0.0074 & 0.0018 \\
			\text{high educated}${}_1$ $\times$  & 0.0122***  &  0.0030***  & 0.0165     & 0.0041   \\
			\text{part-time employed}${}_1$ & 0.0035 & 0.0006 & 0.0106 & 0.0026 \\
			\text{high educated}${}_1$ $\times$  & -0.0248***   & -0.0025*** & -0.0117  & -0.0029     \\
			 \text{part-time employed}${}_5$ & 0.0038 & 0.0005 & 0.0092 & 0.0022 \\
			\text{high educated}${}_1$ $\times$ & 0.1456***  & 0.0120*** & 0.0595***      & 0.0146***   \\
			\text{marginal employed}${}_2$ & 0.0061  & 0.0005 & 0.0083 & 0.0020 \\
			\text{high educated}${}_1$ $\times$  & -0.0809***   & -0.0152***  & -0.0797***     & -0.0196***     	\\
			\text{strong positive wage growth}${}_1$ & 0.0030 & 0.0006 & 0.0103 & 0.0025   \\ 
			\text{high educated}${}_1$ $\times$   &  -0.0481***  & -0.0092***  & -0.0508***    & -0.0125***  	\\
			\text{strong positive wage growth}${}_2$ & 0.0030 & 0.0006 & 0.0102 & 0.0025 \\
			\text{high educated}${}_1$ $\times$   & -0.0371***   & -0.0065***  & -0.0362***   & -0.0089***   	\\ 
			\text{strong positive wage growth}${}_3$ & 0.0032 &  0.0006 &  0.0102 &  0.0025\\
			\text{high educated}${}_1$ $\times$  &  -0.0253***   & -0.0040***  & -0.0244**   &  -0.0060**   	\\ 
			\text{strong positive wage growth}${}_4$ & 0.0033 & 0.0006 & 0.0104 & 0.0025  \\
			\text{high educated}${}_1$ $\times$   &  -0.0307***  & -0.0050***& -0.0280***      & -0.0069*** 	\\
			\text{strong positive wage growth}${}_5$ & 0.0032 &  0.0005  & 0.0098 & 0.0024 \\
			\text{high educated}${}_1$ $\times$  &  -0.1650***  & -0.0419***  & -0.1993***  & -0.0489***	\\ 
			 \text{employed}${}_1$ & 0.0042 & 0.0012 & 0.0211 &  0.0052 \\
			\text{high educated}${}_1$ $\times$  & 0.0369***  & -0.0115***  & -0.0578***  & -0.0142***   \\ 
			\text{change of establishment}${}_1$  &  0.0039 & 0.0006 &    0.0113 &  0.002 \\                                            
			\text{high educated}${}_1$ $\times$  & 0.0028***  & 0.0837***  & 0.4135***  & 0.1014*** 	\\ 
			\text{daily wage}${}_1$ &  0.00004 & 0.0013 & 0.0235 & 0.0057 \\
			\text{high educated}${}_1$ $\times$  & 0.0007***  & 0.0213***  & 0.1099***  & 0.0269***     	\\ 
			\text{daily wage}${}_3$ & 0.00004 & 0.0013 & 0.0234 & 0.0057 \\
			\text{high educated}${}_1$ $\times$  & 0.0001**   & 0.0046***  & 0.0232 &  0.0057      	\\
			 \text{daily wage}${}_4$ & 0.00005 & 0.0015 &   0.0267 & 0.0066 \\
			\text{high educated}${}_1$ $\times$   & 0.0004***   &  0.0144***  & 0.0730***  & 0.0179***  \\
			\text{daily wage}${}_5$ & 0.00004 & 0.0012 & 0.0217 &  0.0053	\\[0,2cm]    
			\multicolumn{5}{l}{$^+$ Variables are standardized.} \\
			\multicolumn{5}{l}{$^{++}$ Only continuous baseline variables are standardized.} \\
			\multicolumn{5}{l}{$^*$, $^{**}$, $^{***}$ statistically significant at the 10\%, 5\% and 1\% levels, respectively.}
		\end{tabular}
	\end{center}
\end{table*}
%

\begin{table*}
	\begin{center}
		\small
		\caption{\small Inference}\label{Tab: 3}
		\begin{tabular}[!htbp]{lSSS}    \\[-0.5em] 
			$\text{LCD}_{CV}$ &    \\[0,2cm]   
			$\boldsymbol{\hat{\mathcal{S}}}_{\text{FDR}}$ & \text{OLS Estimate}$^+$ & \text{Estimate}$^+$  & \text{Marginal effect}$^+$  \\
			& \text{(with refit)} & \text{(with refit)} & \text{(with refit)}  \\
			$\text{daily wage}_1$ & 0.0354***  & 0.1434***  & 0.0355***   \\  
			 & 0.0009 & 0.014 & 0.0034 \\
			$\text{daily wage}_2$ &  0.0187***  & 0.0760***  & 0.0188*** \\
			& 0.0009 & 0.014 &  0.0035 \\
			\multicolumn{4}{l}{$^+$ Variables are standardized.} \\
			\multicolumn{4}{l}{$^*$, $^{**}$, $^{***}$ statistically significant at the 10\%, 5\% and 1\% levels, respectively.}
		\end{tabular}
	\end{center}
\end{table*}

\newpage
\begin{longtable}[!htbp]{lSSS}   \\[-0.5em] 
	\caption{\small Inference}\label{Tab: 4}\\[1 cm]  
	\endfirsthead	
	\caption*{\textbf{Table \ref{Tab: 4}:} Continued}\\
	\endhead
	\endfoot
\small 
			$\text{LSM}$ &    \\[0,2cm]   
			$\boldsymbol{\hat{\mathcal{S}}}_{\text{FDR}}$ & \text{OLS Estimate}$^+$  & \text{Estimate}$^+$  & \text{Marginal effect}$^+$  \\
			& \text{(with refit)} & \text{(with refit)} & \text{(with refit)} \\
			$\text{employed}_1$  & -0.0033***   & -0.0142* & -0.0035* \\
			& 0.0005 &  0.0076 & 0.0019 \\
			$\text{marginal employed}_3$  &  0.0044***  & 0.0191**  & 0.0047**  \\
			& 0.0005 & 0.0078 & 0.0019 \\
			$\text{change of establishment}_1$ & 0.0010**  & 0.0036 & 0.0009   \\
			& 0.0004 &  0.0073 & 0.0018 \\
			$\text{change of establishment}_4$  & 0.0026*** & 0.0106  & 0.0026  \\
			&  0.0004 & 0.0074 & 0.0018 \\
			$\text{change of establishment}_5$  & 0.0025***  & 0.0105  & 0.0026 \\
			& 0.0004 & 0.0073 & 0.0018  \\
			$\text{strong negative wage growth}_5$ & 0.0032*** & 0.0133*  & 0.0033*   \\
			& 0.0004  & 0.0073 & 0.0018 \\
			$\text{strong positive wage growth}_2$ & 0.0103***  & 0.0418**  &  0.0102**    \\
			& 0.0012 & 0.0202 & 0.0050   \\
			$\text{wage profile 1}$ & 0.0012*** &  0.0053  & 0.0013   \\
			& 0.0004  & 0.0074 &  0.0018 \\
			$\text{high educated}_1 \times$   & -0.0045***  &  -0.0208*  & -0.0051  \\
			$\text{part-time employed}_1$ & 0.0006 & 0.0095 & 0.0023 \\
			$\text{high educated}_1 \times$  & -0.0020*** & -0.0089  & -0.0022  \\
			 $\text{part-time employed}_5$ & 0.0005 & 0.0091 & 0.0022 \\
			$\text{high educated}_1 \times$  & 0.0000  & -0.0005  & -0.0001   \\
			  $\text{marginal employed}_1$ & 0.0006 & 0.0104 & 0.0025 \\
			$\text{high educated}_1 \times$  &  0.0039***  & 0.0189  & 0.0046  \\
			$\text{marginal employed}_2$  & 0.0007 & 0.0116 & 0.0028 \\
			$\text{high educated}_1 \times$  & 0.0028***  & 0.0156 & 0.0038 \\
			$\text{marginal employed}_3$  & 0.0006 &  0.0100 &  0.0025 \\
			$\text{high educated}_1 \times$  & -0.0176***  & -0.0905***  & -0.0222***  \\
			$\text{strong positive wage growth}_1$ & 0.0006 & 0.0101 & 0.0025 \\
			$\text{high educated}_1 \times$ & -0.0134***  & -0.0706***  &  -0.0173*** 	\\  
			$\text{strong positive wage growth}_2$ & 0.0006 &  0.0102 & 0.0025 \\
			$\text{high educated}_1 \times$  & -0.0087***  & -0.0476*** &  -0.0117***  \\ 
			$\text{strong positive wage growth}_3$ & 0.0006 & 0.0101  & 0.0025 \\
			$\text{high educated}_1 \times$  & -0.0055***  & -0.0324***  & -0.0079***  \\
			$\text{strong positive wage growth}_4$ & 0.0006 & 0.0103 & 0.0025 \\  
			$\text{high educated}_1 \times$  & -0.0053***  & -0.0301***  & -0.0074*** \\ 
			 $\text{strong positive wage growth}_5$ & 0.0005 &  0.0097 &  0.0024 \\
			$\text{high educated}_1 \times$   & -0.0149***& -0.0731***    & -0.0179***  	\\
			 $\text{change of establishment}_1$ &   0.0006  & 0.0112 & 0.0027 \\
			$\text{high educated}_1 \times$ & 0.0631***   & 0.3063***  & 0.0750***  \\
			$\text{daily wage}_1$  & 0.0011 & 0.0196 & 0.0048  \\
			$\text{high educated}_1 \times$ & 0.0209***  & 0.1104***  & 0.0270***   	\\
			$\text{daily wage}_3$ & 0.0013 & 0.0234 & 0.0057 \\
			$\text{high educated}_1 \times$   & 0.0051*** & 0.0272 &  0.0067    	\\ 
			$\text{daily wage}_4$ &  0.0014 &  0.0266 & 0.0065 \\
			$\text{high educated}_1 \times$   & 0.0029  & 0.0246 &  0.0060	\\ 
			$\text{daily wage}_5$ & 0.0012 &  0.0218 &  0.0054 \\
			$\text{experience}_2 \times$   & -0.0117***  & -0.0470**  & -0.0115**    \\
			$\text{strong positive wage growth}_2$ & 0.0012 &  0.0197 & 0.0048 \\
			$\text{experience}_5 \times$   & 0.0276***  &  0.1143*** &  0.0280***    \\
			$\text{daily wage}_5$ & 0.0005 & 0.0086 &  0.0021  \\[0,2cm]
			\multicolumn{4}{l}{$^+$ Variables are standardized.} \\
			\multicolumn{4}{l}{$^*$, $^{**}$, $^{***}$ statistically significant at the 10\%, 5\% and 1\% levels.}
\end{longtable}
%
%
%
\newpage
\begin{longtable}[!htbp]{p{4.5cm} p{12cm} }  \hline \\[-0.5em] 
	\caption{Legend of Baseline Predictors}\label{Tab:2}\\[0,5cm] \toprule \\[-0.5em]
	\endfirsthead	
	\caption*{\textbf{Table \ref{Tab:2}:} Continued}\\\toprule
	\endhead
	\endfoot
	\bottomrule
	\endlastfoot
	\small
	Variable & Description \\[0,2cm] \hline \\[-0.5em]
	career & 1 if woman makes a professional career, 0 otherwise \\
	age & Women's age in years \\	
	low educated	 & 1 if	no postsecondary degree, 0 otherwise \\ 
	high educated	& 1 if tertiary educated, 0 otherwise  \\
	non-German citizenship & 1 if with migration background, 0 otherwise \\
	employed	& 1 if woman is employed at least 50\% of the year, 0 otherwise \\
	part-time employed & 1 if working part-time, 0 otherwise    \\
	marginally employed & 1 if woman works in a minijob, 0 otherwise \\
	daily wage & Daily wage in Euros   \\
	wage profile 1 & 1 if the wage level anually increases over the past five years, 0 otherwise \\
	wage profile 2 & 1 if the wage increases by 10\% in the last 5 years, 0 otherwise \\
	years since last change of establishment & How many years have passed since the last change of business \\
	change of establishment &  1 if change of business, 0 otherwise \\
	strong negative wage \newline growth &  1 if wage growth $<=-0.05$, 0 otherwise \\
	strong positive wage \newline growth  &  1 if wage growth $>= 0.05$  \\
	experience	& Women's working experience in years \\
	$\text{experience}^2$ & 1 , 0 otherwise \\
	$\text{days of unemployment}$ \newline  $\text{benefits}$ & Unemployment benefit days per year, 0 otherwise \\
	$\text{days of registered job}$ \newline $\text{search while not employed}$ &  Days of being not unemployed and job seeking per year , 0 otherwise \\
	$\text{unemployment benefits}$ &  1 if unemployment benefits, 0 otherwise \\
	$\text{registered job search}$ \newline $\text{while not employed}$ &  1 if not unemployed and job seeking, 0 otherwise \\
	$\text{occupation1}$ &  1 if plant breeders, animal breeders, fishery professions, 0 otherwise \\
	$\text{occupation2}$ &  1 if miners, stone workers, manufacturers of building materials, 0 otherwise \\
	$\text{occupation3}$ &  1 if ceramist, glassmaker, 0 otherwise \\
	$\text{occupation4}$ &  1 if chemical worker, plastic processor, 0 otherwise \\
	$\text{occupation5}$ &  1 if paper manufacturer, printer , 0 otherwise \\
	$\text{occupation6}$ &  1 if woodworker, 0 otherwise \\
	$\text{occupation7}$ &  1 if metal producer, 0 otherwise \\
	$\text{occupation8}$ &  1 if locksmith, mechanic, 0 otherwise \\
	$\text{occupation9}$ &  1 if electrician, 0 otherwise \\
	$\text{occupation10}$ &  1 if assemblers, 0 otherwise \\
	$\text{occupation11}$ &  1 if textiles, and clothing professions, 0 otherwise \\
	$\text{occupation12}$ &  1 if nutrition professionals, 0 otherwise \\
	$\text{occupation13}$ &  1 if building jobs, 0 otherwise \\
	$\text{occupation14}$ &  1 if exterior/interior decorator,  upholsterer, 0 otherwise \\
	$\text{occupation15}$ &  1 if carpenter, modeler, 0 otherwise \\
	$\text{occupation16}$ &  1 if painter, 0 otherwise \\
	$\text{occupation17}$ &  1 if goods inspector, 0 otherwise \\
	$\text{occupation18}$ &  1 if auxiliary worker, 0 otherwise \\
	$\text{occupation19}$ &  1 if machinists, 0 otherwise \\
	$\text{occupation20}$ &  1 if engineers, chemists, physicists, mathematicians, 0 otherwise \\
	$\text{occupation21}$ &  1 if technician, 0 otherwise \\
	$\text{occupation22}$ &  1 if were merchants, 0 otherwise \\
	$\text{occupation23}$ &  1 if services consultants, 0 otherwise  \\
	$\text{occupation24}$ &  1 if traffic professions, 0 otherwise \\
	$\text{occupation25}$ &  1 if organizational or administrative office occupations, 0 otherwise \\
	$\text{occupation26}$ &  1 if regulatory or security occupations, 0 otherwise \\
	$\text{occupation27}$ &  1 if writers and artistic professions, 0 otherwise  \\
	$\text{occupation28}$ &  1 if health care professionals, 0 otherwise \\
	$\text{occupation29}$ &  1 if social and educational occupations, 0 otherwise \\
	$\text{occupation30}$ &  1 if general service occupations, 0 otherwise   \\ \hline \\[-0.5em]		
\end{longtable}	

\newpage
\singlespacing
\textbf{Full list of predictor variables:}
\begin{align*}
\mathcal{A}^{full} =\{&\text{career}_1,\text{career}_2,\text{career}_3,\text{career}_4,\text{career}_5,\text{wage}_1,\text{wage}_2,\text{wage}_3,\text{wage}_4,\text{wage}_5,\\
&\text{full-time employed}_1,\text{full-time employed}_2,\text{full-time employed}_3,\text{full-time employed}_4, \\
&\text{full-time employed}_5,\text{part-time employed}_1, \text{part-time employed}_2, \\
&\text{part-time employed}_3,\text{part-time employed}_4,\text{part-time employed}_5,\\
&\text{marginally employed}_1,\text{marginally employed}_2,\text{marginally employed}_3,\text{marginally employed}_4, \\
&\text{marginally employed}_5,\text{change of establishment}_1,\text{change of establishment}_2, \\
&\text{change of establishment}_3, \text{change of establishment}_4, \text{change of establishment}_5, \\
&\text{years since last change of establishment}_1, \text{years since last change of establishment}_2, \\ 
&\text{years since last change of establishment}_3, \text{years since last change of establishment}_4,        \\
&\text{years since last change of establishment}_5, \text{strong negative wage growth}_1,  \\
&\text{strong negative wage growth}_2,\text{strong negative wage growth}_3,\text{strong negative wage growth}_4,   \\
&\text{strong negative wage growth}_5,\text{strong positive wage growth}_1,\text{strong positive wage growth}_2,    \\
&\text{strong positive wage growth}_3,\text{strong positive wage growth}_4,\text{strong positive wage growth}_5, \\
&\text{experience}_1,\text{experience}^2_1,\text{experience}^2_2,\text{experience}^2_3,\text{experience}^2_4,\text{experience}^2_5, \\
&\text{unemployment benefit days}_1,\text{unemployment benefit days}_2,\text{unemployment benefit days}_3, \\
&\text{unemployment benefit days}_4, \text{unemployment benefit days}_5, \\
&\text{registered job search days while not employed}_1,    \\
&\text{registered job search days while not employed}_2,    \\
&\text{registered job search days while not employed}_3, \text{unemployment benefit}_1,       \\
&\text{unemployment benefit}_2,\text{unemployment benefit}_3,\text{unemployment benefit}_4, \\
&\text{unemployment benefit}_5, \text{registered job search while not employed}_1,                \\
&\text{registered job search while not employed}_2,\text{registered job search while not employed}_3,    \\
&\text{low educated}_1, \text{high educated}_1, \text{non-German citizenship}_1, \text{occupation1}_1    \\
&\text{occupation2}_1,\text{occupation3}_1,\text{occupation4}_1,\text{occupation5}_1,\text{occupation6}_1,\text{occupation7}_1 \\
&\text{occupation8}_1,\text{occupation9}_1,\text{occupation10}_1,\text{occupation11}_1,\text{occupation12}_1,\text{occupation13}_1\\
&\text{occupation14}_1,\text{occupation15}_1,\text{occupation16}_1,\text{occupation17}_1,\text{occupation18}_1,\text{occupation19}_1 \\
&\text{occupation20}_1, \text{occupation21}_1,\text{occupation22}_1,\text{occupation23}_1,\text{occupation24}_1,\text{occupation25}_1 \\
&\text{occupation26}_1,\text{occupation27}_1,\text{occupation28}_1,\text{occupation29}_1,\text{occupation30}_1 \\
&\text{age}_1,\text{age}^2_1,\text{wage profile 1},\text{wage profile 2}    \\
&\text{high educated}_1 \times \text{experience}_1,\text{high educated}_1 \times \text{experience}^2_1,\text{high educated}_1 \times \text{experience}_2   \\
&\text{high educated}_1 \times \text{part-time employed}_1, \text{high educated}_1 \times \text{part-time employed}_2, \text{high educated}_1 \times \text{part-time employed}_3, \\
&\text{high educated}_1 \times \text{part-time employed}_4,\text{high educated}_1 \times \text{part-time employed}_5,    \\
&\text{high educated}_1 \times \text{marginally employed}_1, \text{high educated}_1 \times \text{marginally employed}_2, \\
&\text{high educated}_1 \times \text{marginally employed}_3, \text{high educated}_1 \times \text{marginally employed}_4, \\
&\text{high educated}_1 \times \text{marginally employed}_5, \text{high educated}_1 \times \text{strong negative wage growth}_1, \\
&\text{high educated}_1 \times \text{strong negative wage growth}_2, \\
&\text{high educated}_1 \times \text{strong negative wage growth}_3, \\
\end{align*}

\newpage
\begin{align*}
&\text{high educated}_1 \times \text{strong negative wage growth}_4,  \\
&\text{high educated}_1 \times \text{strong negative wage growth}_5, \text{high educated}_1 \times \text{strong positive wage growth}_1,   \\
&\text{high educated}_1 \times \text{strong positive wage growth}_2, \text{high educated}_1 \times \text{strong positive wage growth}_3, \\
&\text{high educated}_1 \times \text{strong positive wage growth}_4, \text{high educated}_1 \times \text{strong positive wage growth}_5, \\
&\text{high educated}_1 \times \text{full-time employed}_1, \text{high educated}_1 \times \text{full-time employed}_2,  \\
&\text{high educated}_1 \times \text{full-time employed}_3, \text{high educated}_1 \times \text{full-time employed}_4, \\
&\text{high educated}_1 \times \text{full-time employed}_5, \\
&\text{high educated}_1 \times \text{years since last change of establishment}_1, \\
&\text{high educated}_1 \times \text{years since last change of establishment}_2, \\
&\text{high educated}_1 \times \text{years since last change of establishment}_3, \\
&\text{high educated}_1 \times \text{years since last change of establishment}_4, \\
&\text{high educated}_1 \times \text{years since last change of establishment}_5, \\
&\text{high educated}_1 \times \text{change of establishment}_1, \text{high educated}_1 \times \text{change of establishment}_2, \\
&\text{high educated}_1 \times \text{change of establishment}_3, \text{high educated}_1 \times \text{change of establishment}_4, \\
&\text{high educated}_1 \times \text{change of establishment}_5, \text{high educated}_1 \times \text{daily wage}_1, \\
&\text{high educated}_1 \times \text{daily wage}_2, \text{high educated}_1 \times \text{daily wage}_3, \text{high educated}_1 \times \text{daily wage}_4,\\
&\text{high educated}_1 \times \text{daily wage}_5,    \\
&\text{high educated}_1 \times \text{unemployment benefit days}_1, \text{high educated}_1 \times \text{unemployment benefit days}_2, \\
&\text{high educated}_1 \times \text{unemployment benefit days}_3, \text{high educated}_1 \times \text{unemployment benefit days}_4, \\
&\text{high educated}_1 \times \text{unemployment benefit days}_5,  \\
&\text{high educated}_1 \times \text{registered job search days while employed}_1,\\
&\text{high educated}_1 \times \text{registered job search days while employed}_2, \\
&\text{high educated}_1 \times \text{registered job search days while employed}_3, \text{high educated}_1 \times \text{unemployment benefit}_1, \\
&\text{high educated}_1 \times \text{unemployment benefit}_2, \text{high educated}_1 \times \text{unemployment benefit}_3, \\
&\text{high educated}_1 \times \text{unemployment benefit}_4, \text{high educated}_1 \times \text{unemployment benefit}_5, \\
&\text{high educated}_1 \times \text{registered job search days while employed}_1, \\
&\text{high educated}_1 \times \text{registered job search days while employed}_2, \\
&\text{high educated}_1 \times \text{registered job search days while employed}_3, \\
&\text{low educated}_1 \times \text{experience}_1, \text{low educated}_1 \times \text{experience}^2_1, \text{low educated}_1 \times \text{experience}^2_2 \\
&\text{low educated}_1 \times \text{part-time employed}_1, \text{low educated}_1 \times \text{part-time employed}_2, \\
&\text{low educated}_1 \times \text{part-time employed}_3, \text{low educated}_1 \times \text{part-time employed}_4, \\
&\text{low educated}_1 \times \text{part-time employed}_5, \text{low educated}_1 \times \text{marginally employed}_1, \\
&\text{low educated}_1 \times \text{marginally employed}_2, \text{low educated}_1 \times \text{marginally employed}_3, \\
&\text{low educated}_1 \times \text{marginally employed}_4, \text{low educated}_1 \times \text{marginally employed}_5, \\
&\text{low educated}_1 \times \text{strong negative wage growth}_1, \text{low educated}_1 \times \text{strong negative wage growth}_2, \\
&\text{low educated}_1 \times \text{strong negative wage growth}_3, \text{low educated}_1 \times \text{strong negative wage growth}_4, \\
&\text{low educated}_1 \times \text{strong negative wage growth}_5, \text{low educated}_1 \times \text{strong positive wage growth}_1,  \\
&\text{low educated}_1 \times \text{strong positive wage growth}_2, \text{low educated}_1 \times \text{strong positive wage growth}_3, \\
&\text{low educated}_1 \times \text{strong positive wage growth}_4, \text{low educated}_1 \times \text{strong positive wage growth}_5, \\ 
\end{align*}

\newpage 

\begin{align*}
&\text{low educated}_1 \times \text{employed}_1, \text{low educated}_1 \times \text{employed}_2, \text{low educated}_1 \times \text{employed}_3 \\
&\text{low educated}_1 \times \text{employed}_4, \text{low educated}_1 \times \text{employed}_5, \\
&\text{low educated}_1 \times \text{years since last change of establishment}_1, \\
&\text{low educated}_1 \times \text{years since last change of establishment}_2, \\
&\text{low educated}_1 \times \text{years since last change of establishment}_3, \\
&\text{low educated}_1 \times \text{years since last change of establishment}_4, \\
&\text{low educated}_1 \times \text{years since last change of establishment}_5, \\
&\text{low educated}_1 \times \text{change of establishment}_1, \text{low educated}_1 \times \text{change of establishment}_2, \\
&\text{low educated}_1 \times \text{change of establishment}_3, \text{low educated}_1 \times \text{change of establishment}_4, \\
&\text{low educated}_1 \times \text{change of establishment}_5, \text{low educated}_1 \times \text{daily wage}_1, \text{low educated}_1 \times \text{daily wage}_2,\\
&\text{low educated}_1 \times \text{daily wage}_3, \text{low educated}_1 \times \text{daily wage}_4, \text{low educated}_1 \times \text{daily wage}_5, \\
&\text{low educated}_1 \times \text{unemployment benefit days}_1, \text{low educated}_1 \times \text{unemployment benefit days}_2, \\
&\text{low educated}_1 \times \text{unemployment benefit days}_3, \text{low educated}_1 \times \text{unemployment benefit days}_4, \\
&\text{low educated}_1 \times \text{unemployment benefit days}_5,  \\
&\text{low educated}_1 \times \text{registered job search days while employed}_1,\\
&\text{low educated}_1 \times \text{registered job search days while employed}_2, \\
&\text{low educated}_1 \times \text{registered job search days while employed}_3, \text{low educated}_1 \times \text{unemployment benefit}_1, \\
&\text{low educated}_1 \times \text{unemployment benefit}_2, \text{low educated}_1 \times \text{unemployment benefit}_3, \\
&\text{low educated}_1 \times \text{unemployment benefit}_4, \text{low educated}_1 \times \text{unemployment benefit}_5, \\
&\text{low educated}_1 \times \text{registered job search days while employed}_1, \\
&\text{low educated}_1 \times \text{registered job search days while employed}_2, \\
&\text{low educated}_1 \times \text{registered job search days while employed}_3, \\
&\text{high educated}_1 \times \text{wage profile1}_1, \text{high educated}_1 \times \text{wage profile2}_1, \\
&\text{high educated}_1 \times \text{non-German citizenship}_1, \text{low educated}_1 \times \text{wage profile1}_1, \\
& \text{low educated}_1 \times \text{wage profile2}_1, \text{low educated}_1 \times \text{non-German citizenship}_1 \\
&\text{experience}_1 \times \text{part-time employed}_1, \text{experience}_2 \times \text{part-time employed}_2, \\
&\text{experience}_3 \times \text{part-time employed}_3, \text{experience}_4 \times \text{part-time employed}_4, \\
&\text{experience}_5 \times \text{part-time employed}_5, \text{experience}_1 \times \text{marginally employed}_1,  \\
&\text{experience}_2 \times \text{marginally employed}_2, \text{experience}_3 \times \text{marginally employed}_3,  \\
&\text{experience}_4 \times \text{marginally employed}_4, \text{experience}_5 \times \text{marginally employed}_5,  \\
&\text{experience}_1 \times \text{strong positive wage growth}_1, \text{experience}_2 \times \text{strong positive wage growth}_2, \\
&\text{experience}_3 \times \text{strong positive wage growth}_3, \text{experience}_4 \times \text{strong positive wage growth}_4, \\
&\text{experience}_5 \times \text{strong positive wage growth}_5, \text{experience}_1 \times \text{strong negative wage growth}_1 \\
&\text{experience}_2 \times \text{strong negative wage growth}_2, \text{experience}_3 \times \text{strong negative wage growth}_3 \\
&\text{experience}_4 \times \text{strong negative wage growth}_4, \text{experience}_5 \times \text{strong negative wage growth}_5, \\
&\text{experience}_1 \times \text{employed}_1, \text{experience}_2 \times \text{employed}_2, \text{experience}_3 \times \text{employed}_3, \\
&\text{experience}_4 \times \text{employed}_4, \text{experience}_5 \times \text{employed}_5, \\
&\text{experience}_1 \times \text{years since last change of establishment}_1, \\
\end{align*}

\newpage
\begin{align*}
&\text{experience}_2 \times \text{years since last change of establishment}_2, \\
&\text{experience}_3 \times \text{years since last change of establishment}_3, \\
&\text{experience}_4 \times \text{years since last change of establishment}_4, \\
&\text{experience}_5 \times \text{years since last change of establishment}_5, \\
&\text{experience}_1 \times \text{change of establishment}_1, \text{experience}_2 \times \text{change of establishment}_2, \\
&\text{experience}_3 \times \text{change of establishment}_3, \text{experience}_4 \times \text{change of establishment}_4, \\
&\text{experience}_5 \times \text{change of establishment}_5, \text{experience}_1 \times \text{daily wage}_1,  \text{experience}_2 \times \text{daily wage}_2      \\
& \text{experience}_3 \times \text{daily wage}_3, \text{experience}_4 \times \text{daily wage}_4, \text{experience}_5 \times \text{daily wage}_5 \\
&\text{experience}_1 \times \text{unemployment benefit days}_1,\text{experience}_2 \times \text{unemployment benefit days}_2, \\
&\text{experience}_3 \times \text{unemployment benefit days}_3,\text{experience}_4 \times \text{unemployment benefit days}_4, \\
&\text{experience}_5 \times \text{unemployment benefit days}_5, \\
&\text{experience}_1 \times \text{registered job search days while employed}_1,\\
&\text{experience}_2 \times \text{registered job search days while employed}_2, \\
&\text{experience}_3 \times \text{registered job search days while employed}_3,  \\
&\text{experience}_1 \times \text{unemployment benefit}_1, \text{experience}_2 \times \text{unemployment benefit}_2, \\
&\text{experience}_3 \times \text{unemployment benefit}_3, \text{experience}_4 \times \text{unemployment benefit}_4, \\
&\text{experience}_5 \times \text{unemployment benefit}_5, \text{experience}_1 \times \text{registered job search while unemployed}_1, \\
&\text{experience}_2 \times \text{registered job search while unemployed}_2,  \\
&\text{experience}_3 \times \text{registered job search while unemployed}_3, \\
&\text{experience}^2_1 \times \text{part-time employed}_1, \text{experience}^2_2 \times \text{part-time employed}_2, \\
&\text{experience}^2_3 \times \text{part-time employed}_3, \text{experience}^2_4 \times \text{part-time employed}_4, \\
&\text{experience}^2_5 \times \text{part-time employed}_5, \text{experience}^2_1 \times \text{marginally employed}_1,  \\
&\text{experience}^2_2 \times \text{marginally employed}_2, \text{experience}^2_3 \times \text{marginally employed}_3,  \\
&\text{experience}^2_4 \times \text{marginally employed}_4, \text{experience}^2_5 \times \text{marginally employed}_5,  \\
&\text{experience}^2_1 \times \text{strong positive wage growth}_1, \text{experience}^2_2 \times \text{strong positive wage growth}_2, \\
&\text{experience}^2_3 \times \text{strong positive wage growth}_3, \text{experience}^2_4 \times \text{strong positive wage growth}_4, \\
&\text{experience}^2_5 \times \text{strong positive wage growth}_5, \text{experience}^2_1 \times \text{strong negative wage growth}_1 \\
&\text{experience}^2_2 \times \text{strong negative wage growth}_2, \text{experience}^2_3 \times \text{strong negative wage growth}_3 \\
&\text{experience}^2_4 \times \text{strong negative wage growth}_4, \text{experience}^2_5 \times \text{strong negative wage growth}_5, \\
&\text{experience}^2_1 \times \text{employed}_1, \text{experience}^2_2 \times \text{employed}_2, \text{experience}^2_3 \times \text{employed}_3, \\
&\text{experience}^2_4 \times \text{employed}_4, \text{experience}^2_5 \times \text{employed}_5, \\
&\text{experience}^2_1 \times \text{years since last change of establishment}_1, \\
\end{align*}
\newpage
\begin{align*}
&\text{experience}^2_2 \times \text{years since last change of establishment}_2, \\
&\text{experience}^2_3 \times \text{years since last change of establishment}_3, \\
&\text{experience}^2_4 \times \text{years since last change of establishment}_4, \\
&\text{experience}^2_5 \times \text{years since last change of establishment}_5, \\
&\text{experience}^2_1 \times \text{change of establishment}_1, \text{experience}^2_2 \times \text{change of establishment}_2, \\
&\text{experience}^2_3 \times \text{change of establishment}_3, \text{experience}^2_4 \times \text{change of establishment}_4, \\
&\text{experience}^2_5 \times \text{change of establishment}_5, \text{experience}^2_1 \times \text{daily wage}_1,  \text{experience}^2_2 \times \text{daily wage}_2      \\
& \text{experience}^2_3 \times \text{daily wage}_3, \text{experience}^2_4 \times \text{daily wage}_4, \text{experience}^2_5 \times \text{daily wage}_5 \\
&\text{experience}^2_1 \times \text{unemployment benefit days}_1,\text{experience}^2_2 \times \text{unemployment benefit days}_2, \\
&\text{experience}^2_3 \times \text{unemployment benefit days}_3,\text{experience}^2_4 \times \text{unemployment benefit days}_4, \\
&\text{experience}^2_5 \times \text{unemployment benefit days}_5, \\
&\text{experience}^2_1 \times \text{registered job search days while employed}_1,\\
&\text{experience}^2_2 \times \text{registered job search days while employed}_2, \\
&\text{experience}^2_3 \times \text{registered job search days while employed}_3,  \\
&\text{experience}^2_1 \times \text{unemployment benefit}_1, \text{experience}^2_2 \times \text{unemployment benefit}_2, \\
&\text{experience}^2_3 \times \text{unemployment benefit}_3, \text{experience}^2_4 \times \text{unemployment benefit}_4, \\
&\text{experience}^2_5 \times \text{unemployment benefit}_5, \text{experience}^2_1 \times \text{registered job search while unemployed}_1, \\
&\text{experience}^2_2 \times \text{registered job search while unemployed}_2,  \\
&\text{experience}^2_3 \times \text{registered job search while unemployed}_3, \\
&\text{experience}_1 \times \text{wage profile 1}, \text{experience}_1 \times \text{wage profile 2}, \text{experience}^2_1 \times \text{wage profile 1},  \\
&\text{age}_1 \times \text{high educated}_1, \text{age}^2_1 \times \text{high educated}_1, \text{age}_1 \times \text{low educated}_1, \text{age}^2_1 \times \text{low educated}_1.\}
\end{align*}

\end{document}